\documentclass[12pt]{article}%
\usepackage{graphicx}
\usepackage{amssymb}
\usepackage{amsfonts}
\usepackage{amsmath}
\usepackage{graphicx}
\usepackage{lineno}
\usepackage{hyperref}%
\providecommand{\U}[1]{\protect\rule{.1in}{.1in}}
\newtheorem{theorem}{Theorem}[section]
\newtheorem{corollary}{Corollary}[section]

\newtheorem{lemma}{Lemma}[section]
\newtheorem{example}{Example}[section]

\makeatletter
\renewcommand{\@cite}[1]{#1}
\makeatother
\begin{document}

\title{\textbf{A Bias-reduced Estimator for the Mean of a Heavy-tailed Distribution
with an Infinite Second Moment}}
\author{\textbf{Brahim Brahimi, Djamel Meraghni,}
\and \textbf{Abdelhakim Necir\thanks{{\small Corresponding author: (A.~Necir)
\texttt{necirabdelhakim@yahoo.fr}}}, Djabrane Yahia }\medskip\\Laboratory of Applied Mathematics,\\Mohamed Khider University, Biskra, Algeria}
\maketitle

\begin{abstract}
\noindent We use bias-reduced estimators of high quantiles, of heavy-tailed
distributions, to introduce a new estimator of the mean in the case of
infinite second moment. The asymptotic normality of the proposed estimator is
established and checked, in a simulation study, by four of the most popular
goodness-of-fit tests for different sample sizes. Moreover, we compare, in
terms of bias and mean squared error, our estimator with Peng's estimator
(Peng, 2001) and we evaluate the accuracy of some resulting confidence
intervals.\medskip


\end{abstract}

\section{\textbf{Introduction\label{sec1}}}

\noindent Let $X_{1},X_{2},...$ be independent and identically distributed
(i.i.d.) non-negative random variables (r.v.'s) with mean $\mu<\infty,$
variance $\sigma^{2}$ and cumulative distribution function (cdf) $F.\ $Suppose
that the tail of $F$ is regularly varying at infinity with tail index $\left(
-\alpha\right)  <0,$ that is%
\begin{equation}
\lim_{t\rightarrow\infty}\frac{1-F\left(  tx\right)  }{1-F\left(  t\right)
}=x^{-\alpha},\text{ \ \ \ for any }x>0, \label{F}%
\end{equation}
(see, e.g., \cite[, page 19]{deHFe06}). Such cdf's constitute a major subclass
of the family of heavy-tailed distributions. It includes distributions such as
Pareto, Burr, Student, $\alpha-$stable $\left(  0<\alpha<2\right)  ,$ and
log--gamma, which are known to be appropriate models for fitting large
insurance claims, large fluctuations of prices, log--returns, etc. (see, e.g.
\cite{ReTo07}; \cite{BeMaDi01}; \cite{RoScSc99}). In this paper, we are
concerned with the construction of a bias-reduced asymptotically normal
estimator for the mean%
\[
\mu:=\int_{0}^{\infty}xdF\left(  x\right)  ,
\]
which could be rewritten, in terms of the quantile function (corresponding to
the cdf $F)$%
\[
Q(s):=\inf\left\{  x:F\left(  x\right)  \geq s\right\}  ,\text{ }0<s<1,
\]
as%
\begin{equation}
\mu=\int_{0}^{1}Q\left(  1-s\right)  ds. \label{mu}%
\end{equation}
For a given sample $X_{1},...,X_{n},$ let%
\[
{\normalsize Q_{n}\left(  s\right)  :=\inf\left\{  x\in\mathbb{R}:F_{n}\left(
x\right)  \geq s\right\}  ,\;0<s\leq1,}%
\]
denote the sample quantile function (classical non-parametric estimator of
$Q)\ $associated to the empirical cdf defined on the real line by
${\normalsize F_{n}\left(  x\right)  :=n}^{-1}\sum\nolimits_{i=1}%
^{n}\mathbb{I}\left(  X_{i}\leq x\right)  ,${\normalsize \quad}%
with\ $\mathbb{I}\left(  \cdot\right)  $\ being\ the\ indicator\ function. The
natural (unbiased) estimator of $\mu$ is the sample mean%
\begin{equation}
\int_{0}^{1}Q_{n}\left(  1-s\right)  ds=\dfrac{1}{n}%
{\displaystyle\sum\limits_{i=1}^{n}}
X_{i}=:\overline{X}_{n}\text{.} \label{samplemean}%
\end{equation}
From the Central Limit Theorem (CLT), the sequence of r.v.'s $\left\{
\sqrt{n}\left(  \overline{X}_{n}-\mu\right)  /\sigma,\text{ }n\geq1\right\}  $
converges in distribution to the standard Gaussian r.v., provided that the
second-order moment $\mathbf{E}\left[  X_{1}^{2}\right]  $ is finite. This is
a very restrictive condition in the context of heavy-tailed distributions as
the following considerations show. Assume that the r.v. $X_{1}$ follows the
Pareto law with index $\alpha>0,$ that is, $1-F(x)=x^{-\alpha}$ for $x\geq1$.
When $\alpha>1,$ the mean $\mu$ exists, but $\mathbf{E}\left[  X_{1}%
^{2}\right]  $ is only finite for $\alpha\geq2.$ Hence, the range $\alpha
\in(1,2)$ is not covered by the CLT and thus we need to seek another approach
to handle this situation. Making use of Weissman's estimator of high quantiles
\cite{Wei78}, \cite{Pe01} proposed an alternative estimator for $\mu$ and
established its asymptotic normality for any $\alpha\in(1,2).$ Let us define
the following estimator for $Q:$%
\[
\widehat{Q}_{n}(1-s):=\left\{
\begin{array}
[c]{lcc}%
Q_{n}^{W}(1-s) & \text{for } & 0<s<k/n\medskip\\
Q_{n}(1-s) & \text{for} & k/n\leq s<1,
\end{array}
\right.
\]
where%
\begin{equation}
Q_{n}^{W}(1-s):=(k/n)^{1/\widehat{\alpha}_{n}^{H}}X_{n-k,n}s^{-1/\widehat
{\alpha}_{n}^{H}},\text{ }s\downarrow0 \label{qw-formula}%
\end{equation}
is Weissman's estimator of high quantiles, with%
\begin{equation}
\widehat{\alpha}_{n}^{H}:=\left(  k^{-1}\sum\limits_{i=1}^{k}\log
X_{n-i+1,n}-\log X_{n-k,n}\right)  ^{-1}, \label{hill}%
\end{equation}
being the well-known Hill estimator \cite{Hill75} of the tail index $\alpha,$
and $X_{1,n}\leq...\leq X_{n,n}$\ denoting the order statistics pertaining to
the sample $X_{1},...,X_{n}.$ The number $k$ represents the number of upper
order statistics used in the computation of $\widehat{\alpha}_{n}^{H},$ it is
an integer sequence $k=k_{n}$ satisfying%
\begin{equation}
1<k<n,\text{ }k\rightarrow\infty\text{ and }k/n\rightarrow0\text{ as
}n\rightarrow\infty. \label{K}%
\end{equation}
\medskip By replacing $Q\left(  1-s\right)  $ by $\widehat{Q}_{n}(1-t)$ in
formula (\ref{mu}), \cite{Pe01} proposed an alternative estimator for $\mu$ as
follows:%
\[
\widehat{\mu}_{n}^{P}=\widehat{\mu}_{n}^{P}\left(  k\right)  :=\int_{0}%
^{1}\widehat{Q}_{n}(1-s)ds=\int_{0}^{k/n}Q_{n}^{W}\left(  1-s\right)
ds+\int_{k/n}^{1}Q_{n}(1-s)ds,
\]
which, by a straightforward calculation, is equal to%
\begin{equation}
\widehat{\mu}_{n}^{P}:=\frac{k}{n}\frac{\widehat{\alpha}_{n}^{H}}%
{\widehat{\alpha}_{n}^{H}-1}X_{n-k,n}+\frac{1}{n}\sum_{i=k+1}^{n}X_{n-i+1,n},
\label{Peng}%
\end{equation}
provided that $\widehat{\alpha}_{n}^{H}>1.$ Moreover, the same author showed
that, under suitable regularity assumptions, for any $\alpha\in(1,2),$%
\begin{equation}
\frac{\sqrt{n}\left(  \widehat{\mu}_{n}^{P}-\mu\right)  }{\sqrt{k/n}X_{n-k,n}%
}\overset{d}{\rightarrow}\mathcal{N}\left(  0,\sigma^{2}\left(  \alpha\right)
\right)  ,\text{ as }n\rightarrow\infty, \label{normpeng}%
\end{equation}
where%
\[
\sigma^{2}\left(  \alpha\right)  :=\alpha/\left(  1-\alpha\right)  ^{4}\left(
2-\alpha\right)  .
\]
Throughout this paper, the standard notations $\overset{p}{\rightarrow},$
$\overset{d}{\rightarrow}$ and $\overset{d}{=}$ respectively stand for
convergence in probability, convergence in distribution and equality in
distribution, while $\mathcal{N}\left(  a,b^{2}\right)  $\ denotes the normal
distribution with mean $a$ and variance $b^{2}.\medskip$

\noindent Actually, \cite{Pe01} defined his estimator in the more general
situation where the r.v. $X$ is real (not necessarily non-negative) with lower
and upper heavy tails. He simultaneously took into account the regular
variations of both tails of $G$ and the balance condition%
\[
\lim_{t\rightarrow\infty}\left(  1-F\left(  t\right)  \right)  /\left(
1-F\left(  t\right)  -F\left(  -t\right)  \right)  =p\in\left[  0,1\right]  .
\]
In this paper, we only consider non-negative r.v.'s. Our motivation comes from
the actuarial risk theory where insurance losses are represented by such
r.v.'s. In this case, $\widehat{\mu}_{n}^{P}$ may be interpreted as an
estimator of a risk measure called the net premium, see for instance
\cite{NeMe09} and \cite{BrMeNe11}. Note that in our case, since r.v. $X$ is
non-negative, we have $F\left(  -x\right)  =0$ for $x\geq0,$ which yields
$p=1$ in the above balance condition.\medskip

\noindent Hill's estimator $\widehat{\alpha}_{n}^{H}$ plays a pivotal role in
statistical inference on distribution tails. This estimator has been
thoroughly studied, improved and even generalized to any real parameter
$\alpha.$ Weak consistency of\ $\widehat{\alpha}_{n}^{H}$ was established by
\cite{Ma82} assuming only that the underlying cdf $F$ satisfies condition
(\ref{F}). The asymptotic normality of $\widehat{\alpha}_{n}^{H}$ has been
established (see \cite{deHP98}) under the following stricter condition that
characterizes Hall's model (see \cite{Ha82} and \cite{HaWe85}).%
\begin{equation}
1-F\left(  x\right)  =cx^{-\alpha}+dx^{-\beta}+o\left(  x^{-\beta}\right)
,\text{ as }x\rightarrow\infty, \label{A2}%
\end{equation}
for some $c>0,$ $d\neq0$ and $\beta>\alpha>0.$ Note that (\ref{A2}), which is
a special case of a more general second-order regular variation condition (see
\cite{deHSt96}), is equivalent to%
\begin{equation}
Q\left(  1-s\right)  =c^{1/\alpha}s^{-1/\alpha}\left(  1+\alpha^{-1}%
c^{-\beta/\alpha}ds^{\beta/\alpha-1}+o\left(  1\right)  \right)  ,\text{ as
}s\downarrow0. \label{A3}%
\end{equation}
The constants $\alpha$ and $\beta$ are called, respectively, first-order (tail
index, shape parameter) and second-order parameters of cdf $F.$\medskip

\noindent Extreme value based estimators essentially rely on the number $k$ of
upper order statistics involved in estimate computation. Hill's estimator has,
in general, a substantial variance for small values of $k$ and a considerable
bias for large values of $k.$ Hence,\ one has to look for a $k$ value, denoted
by $k^{\ast},$ that balances between these two vices. The choice of this
optimal value $k^{\ast}$ represents a thorny issue in the process of
estimating the tail index and related quantities. To solve this problem,
several adaptive procedures are available, see, e.g., \cite{DedeH93},
\cite{DrKa98}, \cite{DadePd01}, \cite{Ch Pe01}, \cite{NaFr04}, and the
references therein. A theoretical optimal choice of $k$ is obtained by
minimizing the asymptotic mean squared error (RMSE) of $\widehat{\alpha}%
_{n}^{H}.$ Indeed, under condition (\ref{A2}), we have (see \cite{deHP98})%
\begin{equation}
k^{\ast}:=\left(  2^{-1}\alpha\beta^{2}\left(  \beta-\alpha\right)
^{-3}d^{-2}c^{2\beta/\alpha}\right)  ^{\frac{\alpha}{2\beta-\alpha}}%
n^{\frac{2\beta-2\alpha}{2\beta-\alpha}}. \label{kopt}%
\end{equation}
Though Peng's estimator $\widehat{\mu}_{n}^{P}$ enjoys the asymptotic
normality property, it still has a problem due to the fact that, it is based
on Weissman's estimator $Q_{n}^{W}$ known to be largely biased. Fortunately,
many estimators with reduced biases are proposed in the literature as an
alternative to $Q_{n}^{W},$ see, for instance, \cite{FeHa99}, \cite{BeDiSt02},
\cite{GoMa02,GoMa04}, \cite{CaFiGo04,CaGoRo09}, \cite{PeQi04},
\cite{MaDeGuBe04}, \cite{GoFi06}, \cite{GP07} and \cite{BeFiGV08}.\medskip

\noindent In this paper, we use the bias-reduced estimator of the high
quantile $Q\left(  1-s\right)  ,$ recently proposed by \cite{LiPeYa10} who
exploited the censored maximum likelihood (CML) based estimators $\left(
\widehat{\alpha},\widehat{\beta}\right)  $ of the couple of regular variation
parameters $\left(  \alpha,\beta\right)  $ introduced by \cite{PeQi04}. The
CML estimators $\left(  \widehat{\alpha},\widehat{\beta}\right)  $ are defined
as the solution of the two equations (under the constraint $\beta
>\widehat{\alpha}_{n}^{H})$%
\begin{equation}
\frac{1}{k}\sum_{i=1}^{k}\frac{1}{G_{i}\left(  \alpha,\beta\right)  }%
=1\quad\text{and}\quad\frac{1}{k}\sum_{i=1}^{k}\frac{1}{G_{i}\left(
\alpha,\beta\right)  }\log\frac{X_{n-i+1,n}}{X_{n-k,n}}=\beta^{-1},
\label{alpha-beta}%
\end{equation}
where%
\begin{equation}
G_{i}\left(  \alpha,\beta\right)  =\frac{\alpha}{\beta}\left(  1+\frac
{\alpha\beta}{\alpha-\beta}H\left(  \alpha\right)  \right)  \left(
\frac{X_{n-i+1,n}}{X_{n-k,n}}\right)  ^{\beta-\alpha}-\frac{\alpha\beta
}{\alpha-\beta}H\left(  \alpha\right)  , \label{GI}%
\end{equation}
with%
\[
H\left(  \alpha\right)  =\frac{1}{\alpha}-\frac{1}{k}\sum_{i=1}^{k}\log
\dfrac{X_{n-i+1,n}}{X_{n-k,n}}.
\]
\cite{LiPeYa10} obtained their bias-reduced estimators $Q_{n}^{LPY}\left(
1-s\right)  ,$ of the high quantiles $Q\left(  1-s\right)  ,$ by substituting
$\left(  \widehat{\alpha},\widehat{\beta}\right)  $ to $\left(  \alpha
,\beta\right)  $ in (\ref{A3}). That is%
\begin{equation}
Q_{n}^{LPY}\left(  1-s\right)  :=\widehat{c}^{1/\widehat{\alpha}%
}s^{-1/\widehat{\alpha}}\left(  1+\widehat{\alpha}^{-1}\widehat{c}%
^{-\widehat{\beta}/\widehat{\alpha}}\widehat{d}s^{\widehat{\beta}%
/\widehat{\alpha}-1}\right)  ,\text{ }s\downarrow0, \label{LPY}%
\end{equation}
where%
\begin{equation}
\left\{
\begin{array}
[c]{l}%
\widehat{c}=\dfrac{\widehat{\alpha}\widehat{\beta}}{\widehat{\alpha}%
-\widehat{\beta}}\dfrac{k}{n}X_{n-k,n}^{\widehat{\alpha}}\left(  \dfrac
{1}{\widehat{\beta}}-\dfrac{1}{k}%
{\displaystyle\sum\limits_{i=1}^{k}}
\log\dfrac{X_{n-i+1,n}}{X_{n-k,n}}\right)  ,\bigskip\\
\widehat{d}=\dfrac{\widehat{\alpha}\widehat{\beta}}{\widehat{\beta}%
-\widehat{\alpha}}\dfrac{k}{n}X_{n-k,n}^{\widehat{\beta}}\left(  \dfrac
{1}{\widehat{\alpha}}-\dfrac{1}{k}%
{\displaystyle\sum\limits_{i=1}^{k}}
\log\dfrac{X_{n-i+1,n}}{X_{n-k,n}}\right)  .
\end{array}
\right.  \label{c-d}%
\end{equation}
The consistency and asymptotic normality of $Q_{n}^{LPY}\left(  1-s\right)  $
are established by the same authors. Now we can define another estimator for
the quantile function $Q$ as follows:%
\[
\widetilde{Q}_{n}(1-s)=\left\{
\begin{array}
[c]{lcc}%
Q_{n}^{LPY}\left(  1-s\right)  & \text{for} & 0<s<k/n\medskip\\
Q_{n}(1-s) & \text{for} & k/n\leq s<1.
\end{array}
\right.
\]
By replacing $Q$ by $\widetilde{Q}_{n},$ in formula (\ref{mu}), we get%
\begin{equation}
\widehat{\mu}_{n}=\widehat{\mu}_{n}\left(  k\right)  :=\int_{0}^{1}%
\widetilde{Q}_{n}(1-s)ds=\int_{0}^{k/n}Q_{n}^{LPY}\left(  1-s\right)
ds+\int_{k/n}^{1}Q_{n}(1-s)ds. \label{sum}%
\end{equation}
An elementary integral calculation leads to a new bias-reduced estimator for
$\mu$ defined by the following formula:%
\begin{equation}
\widehat{\mu}_{n}:=\left(  k/n\right)  \left(  n\widehat{c}/k\right)
^{1/\widehat{\alpha}}\left(  \frac{\widehat{\alpha}}{\widehat{\alpha}-1}%
+\frac{\widehat{d}\widehat{c}^{-\widehat{\beta}/\widehat{\alpha}}\left(
k/n\right)  ^{\widehat{\beta}/\widehat{\alpha}-1}}{\widehat{\beta}-1}\right)
+\frac{1}{n}\sum_{i=k+1}^{n}X_{n-i+1,n}, \label{newestim}%
\end{equation}
provided that $\widehat{\beta}>\widehat{\alpha}>1$ so that $\widehat{\mu}_{n}$
be finite.\medskip

\noindent The rest of the paper is organized as follows. In Section
\ref{sec2}, we briefly discuss the third order-condition of regular variation
before establishing the asymptotic normality of $\widehat{\mu}_{n}.$\ In
Section \ref{Simulation}, we carry out a simulation study to illustrate the
performance of our new estimator $\widehat{\mu}_{n}$ and compare it with
Peng's one. Proofs are relegated to Section \ref{secProof}. Some concluding
remarks notes made in Section \ref{concluding notess}. Finally, some of the
main results used in Section \ref{secProof} are gathered in the Appendix, as
well as a very brief description of the algorithm of Reiss and Thomas applied,
in Section \ref{Simulation}, to select the optimal sample fraction $k.$

\section{\textbf{Main results\label{sec2}}}

\noindent In the theory of extremes, a function, denoted by $U$ and
(sometimes) called tail quantile function,\ is used quite often. It is defined
by%
\[
U\left(  t\right)  :=\left(  1/\left(  1-F\right)  \right)  ^{-1}\left(
t\right)  =Q\left(  1-1/t\right)  ,\text{ }1<t<\infty.
\]
In terms of this function, Hall's conditions (\ref{A2}) and (\ref{A3}) are
equivalent to%
\begin{equation}
U\left(  t\right)  =c^{1/\alpha}t^{1/\alpha}\left(  1+\alpha^{-1}%
c^{-\beta/\alpha}dt^{1-\beta/\alpha}+o\left(  1\right)  \right)  ,\text{
}t\rightarrow\infty. \label{U}%
\end{equation}
This implies that%
\begin{equation}
\underset{t\rightarrow\infty}{\lim}\dfrac{\log\left[  U\left(  tx\right)
/U\left(  t\right)  \right]  -\alpha^{-1}\log x}{A_{1}\left(  t\right)
}=\dfrac{x^{1-\beta/\alpha}-1}{1-\beta/\alpha},\text{ for any }x>0,
\label{second-order}%
\end{equation}
where%
\[
A_{1}\left(  t\right)  :=d\alpha^{-1}\left(  1-\beta/\alpha\right)
c^{-\beta/\alpha}t^{1-\beta/\alpha}.
\]
The function $A_{1}\left(  t\right)  ,$ which tends to zero as $t\rightarrow
\infty$ (because $\beta>\alpha),$ determines the rate of convergence of
$\log\left[  U\left(  tx\right)  /U\left(  t\right)  \right]  $ to its limit
$\alpha^{-1}\log x.$ Relation (\ref{second-order}) is known as the
second-order condition of regular variation (see, e.g., \cite[page
43]{deHFe06}).\medskip

\noindent Unfortunately, the second-order regular variation\ is not sufficient
to find asymptotic distributions for the estimators defined by the systems
(\ref{alpha-beta}) and (\ref{c-d}). We strengthen it into a condition, called
third-order condition of regular variation and given by (\ref{Third-order}),
that specifies the rate of (\ref{second-order}) (see, e.g., \cite{deHSt96} or
\cite{FrGodeN07}).%
\begin{equation}
\underset{t\rightarrow\infty}{\lim}\frac{\dfrac{\log\left[  U\left(
tx\right)  /U\left(  t\right)  \right]  -\alpha^{-1}\log x}{A_{1}\left(
t\right)  }-\dfrac{x^{1-\beta/\alpha}-1}{1-\beta/\alpha}}{A_{2}\left(
t\right)  }=D\left(  \alpha,\beta,\rho\right)  , \label{Third-order}%
\end{equation}
where $A_{2}\left(  t\right)  \rightarrow0$ as $t\rightarrow\infty,$ with
constant sign near infinity and%
\[
D\left(  \alpha,\beta,\rho\right)  :=\frac{1}{\rho}\left(  \dfrac
{x^{1-\beta/\alpha+\rho}-1}{1-\beta/\alpha+\rho}-\dfrac{x^{1-\beta/\alpha}%
-1}{1-\beta/\alpha}\right)  ,\bigskip
\]
with $\rho$ being a positive constant called third-order parameter.
\cite{PeQi04} established the asymptotic normality of $\widehat{\alpha},$
$\widehat{\beta}$ and $\widehat{c}$ under the following extract conditions on
the sample fraction $k,$ as $n\rightarrow\infty,$%
\begin{equation}
\left(  i\right)  \text{ }\sqrt{k}\left\vert A_{1}\left(  n/k\right)
\right\vert \rightarrow\infty,\text{ \ }\left(  ii\right)  \text{\ }\sqrt
{k}A_{1}^{2}\left(  n/k\right)  \rightarrow0,\text{ \ }\left(  iii\right)
\text{\ }\sqrt{k}A_{1}\left(  n/k\right)  A_{2}\left(  n/k\right)
\rightarrow0. \label{A1A2}%
\end{equation}
As for $\widehat{d},$ it is asymptotically normal under the assumption
$\left(  \sqrt{k}\left\vert A_{1}\left(  n/k\right)  \right\vert \right)
/\log\left(  n/k\right)  \rightarrow\infty$ added to $\left(  ii\right)  $ and
$\left(  iii\right)  .$

\begin{example}
\label{example}Consider the Fr\'{e}chet cdf with shape parameter $\alpha>0$%
\begin{equation}
F\left(  x\right)  =\exp\left(  -x^{-\alpha}\right)  ,\text{ }x>0.
\label{Frechet}%
\end{equation}
The corresponding tail quantile function is defined by $U\left(  t\right)
=(-\log(1-1/t))^{-1\alpha},$ for $t>1.\ $Applying Taylor's expansion (to the
third order) to $U$ and identifying with (\ref{U}), yield $\beta=2\alpha,$
$c=1$ and $d=-1/2.\ $The condition (\ref{Third-order}) holds for $A_{1}\left(
t\right)  =t^{-1}/2\alpha,$ $A_{2}\left(  t\right)  =\left(  \alpha-3\right)
t^{-2}/12\alpha^{2}$ and $\rho=3\alpha.$ Other examples may be found in the
recent paper \cite{GdeW11}.\ The Fr\'{e}chet cdf will be employed, in Section
\ref{Simulation}, as a model in our simulation study.\medskip
\end{example}

\noindent Note that, from a theoretical point of view, assumptions (\ref{K})
and (\ref{A1A2}) are realistic, as the following example shows. Indeed, let us
choose $k=\left[  n^{\epsilon}\right]  ,$ $0<\epsilon<1,$ then it easy to
verify that these assumptions hold for any $2/3<\epsilon<4/5.$ The notation
$\left[  \cdot\right]  $ stands for the integer part of real numbers.\medskip

\noindent Our main result, namely the asymptotic normality of the bias-reduced
estimator $\widehat{\mu}_{n},$ is formulated in the last of the following four
theorems. In Theorem \ref{Theorem1}, we give an approximation of
$\widehat{\alpha}$ in terms of Brownian bridges, which leads to its asymptotic
normality stated in Theorem \ref{Theorem2}. We do the same thing to
$\widehat{\mu}_{n}$ in Theorem \ref{Theorem3}. It is worth mentioning that the
asymptotic normality of $\widehat{\alpha}$ was first established by
\cite{PeQi04}. But, this does not meet our needs to achieve the major object
of this paper. Then, we need to approximate both $\widehat{\alpha}$ and
$\widehat{\mu}_{n}$ by linear functional of the same sequence of standard
Brownian bridges $B_{n}\left(  s\right)  .$

\begin{theorem}
\label{Theorem1}Assume that the third order condition (\ref{Third-order})
holds with $\beta/\alpha=:\lambda>1$ and let $k=k_{n}$ be an integer sequence
satisfying (\ref{K}) and (\ref{A1A2}). Then\ there exists a sequence of
Brownian bridges $\left\{  B_{n}\left(  s\right)  ,\text{ }0\leq
s\leq1\right\}  $\ such that%
\[
\sqrt{k}\left(  \widehat{\alpha}-\alpha\right)  =\alpha\left(  \eta_{1}%
W_{1n}+\eta_{2}W_{2n}+\eta_{3}W_{3n}\right)  +o_{p}\left(  1\right)  ,\text{
as }n\rightarrow\infty
\]
where $W_{1n},$ $W_{2n}$ and $W_{3n}$ are sequences of centered Gaussian
r.v.'s defined by%
\begin{align*}
&  W_{1n}%
\begin{array}
[c]{c}%
:=
\end{array}
\sqrt{n/k}B_{n}\left(  1-k/n\right)  -\sqrt{n/k}%
{\displaystyle\int_{0}^{1}}
s^{-1}B_{n}\left(  1-ks/n\right)  ds,\\
&  W_{2n}%
\begin{array}
[c]{c}%
:=
\end{array}
\left(  \lambda^{-1}-1\right)  \sqrt{n/k}B_{n}\left(  1-k/n\right)  +\left(
\lambda-1\right)  \sqrt{n/k}%
{\displaystyle\int_{0}^{1}}
s^{\lambda-2}B_{n}\left(  1-ks/n\right)  ds,\\
&  W_{3n}%
\begin{array}
[c]{c}%
:=
\end{array}
\left(  1-\lambda\right)  \sqrt{n/k}%
{\displaystyle\int_{0}^{1}}
s^{\lambda-2}\left(  \log s\right)  B_{n}\left(  1-ks/n\right)  ds\\
&  \ \ \ \ \ \ \ \ \ \ \ \ \ \ +\lambda^{-2}\sqrt{n/k}B_{n}\left(
1-k/n\right)  -\sqrt{n/k}%
{\displaystyle\int_{0}^{1}}
s^{\lambda-2}B_{n}\left(  1-ks/n\right)  ds,
\end{align*}
and%
\[
\eta_{1}:=\dfrac{\lambda^{4}}{\left(  \lambda-1\right)  ^{4}},\text{ }\eta
_{2}:=\dfrac{\lambda^{2}\left(  2\lambda-1\right)  \left(  3\lambda-1\right)
}{\left(  \lambda-1\right)  ^{5}}\text{ }\eta_{3}:=\dfrac{\lambda^{3}\left(
2\lambda-1\right)  }{\left(  \lambda-1\right)  ^{4}}^{2}.
\]

\end{theorem}

\begin{theorem}
\label{Theorem2}Under the assumptions of Theorem \ref{Theorem1}, we have%
\begin{equation}
\sqrt{k}\left(  \widehat{\alpha}-\alpha\right)  \overset{d}{\rightarrow
}\mathcal{N}\left(  0,\alpha^{2}\beta^{4}/\left(  \alpha-\beta\right)
^{4}\right)  ,\text{ as }n\rightarrow\infty. \label{asympt-alpha}%
\end{equation}

\end{theorem}

\begin{theorem}
\label{Theorem3}Under the assumptions of Theorem \ref{Theorem1}, we have, as
$n\rightarrow\infty$%
\[
\frac{\sqrt{n}}{\sqrt{k/n}\left(  nc/k\right)  ^{1/\alpha}}\left\{
\widehat{\mu}_{n}-\mu\right\}  =-\frac{\alpha}{\left(  \alpha-1\right)  ^{2}%
}\left\{  \eta_{1}W_{1n}+\eta_{2}W_{2n}+\eta_{3}W_{3n}\right\}  +W_{4n}%
+o_{p}\left(  1\right)  ,\medskip
\]
where $W_{1n},$ $W_{2n}$ and $W_{3n}$ are those of Theorem \ref{Theorem1} and%
\[
W_{4n}:=-\frac{\int_{k/n}^{1}B_{n}\left(  1-s\right)  dQ\left(  1-s\right)
}{\sqrt{k/n}\left(  nc/k\right)  ^{1/\alpha}}.
\]

\end{theorem}

\begin{theorem}
\label{Theorem4}Under the assumptions of Theorem \ref{Theorem1}, we have%
\begin{equation}
\frac{\sqrt{n}}{\sqrt{k/n}\left(  nc/k\right)  ^{1/\alpha}}\left\{
\widehat{\mu}_{n}-\mu\right\}  \overset{d}{\rightarrow}\mathcal{N}\left(
0,\sigma^{2}\left(  \alpha,\beta\right)  \right)  ,\text{ as }n\rightarrow
\infty, \label{asympt-mu}%
\end{equation}
where%
\begin{equation}
\sigma^{2}\left(  \alpha,\beta\right)  :=\frac{\alpha^{2}\beta^{4}}{\left(
\alpha-1\right)  ^{4}\left(  \alpha-\beta\right)  ^{4}}+\frac{2}{2-\alpha
}+\frac{2\alpha\beta^{2}}{\left(  \alpha-1\right)  ^{2}\left(  \alpha
-\beta\right)  ^{2}}. \label{var}%
\end{equation}

\end{theorem}

\noindent The following corollary to Theorem \ref{Theorem4} provides a
straightforward practical way to build confidence intervals for $\mu.$

\begin{corollary}
\label{corollary}Under the assumptions of Theorem \ref{Theorem1}, we have
\[
\frac{\sqrt{n}}{\sqrt{k/n}\sigma\left(  \widehat{\alpha},\widehat{\beta
}\right)  \left(  n\widehat{c}/k\right)  ^{1/\widehat{\alpha}}}\left\{
\widehat{\mu}_{n}-\mu\right\}  \overset{d}{\rightarrow}\mathcal{N}\left(
0,1\right)  ,\text{ as }n\rightarrow\infty,
\]
where $\widehat{\alpha},$ $\widehat{\beta}$ and $\widehat{c}$ are the
estimates of $\alpha,\beta$ and $c$ given in (\ref{alpha-beta}) and
(\ref{c-d}) respectively.
\end{corollary}

\section{\textbf{Illustrative simulation study\label{Simulation}}}

\noindent Let $z_{\zeta}$ denote $\left(  1-\zeta/2\right)  $-quantile of the
standard normal r.v. Given a realization $\left(  x_{1},...,x_{n}\right)  $ of
$\left(  X_{1},...,X_{n}\right)  $ from a population$X$ satisfying the
required assumptions, we construct a $\left(  1-\zeta/2\right)  100\%$
confidence interval for $\mu$ via the following four steps:\medskip

\noindent\textbf{Step 1:} Applying Reiss and Thomas algorithm (see subsection
\ref{RT} of the Appendix), we select the optimal sample fraction $k^{\ast}.$

\noindent\textbf{Step 2:} Resolving the system (\ref{alpha-beta}) with
$k=k^{\ast},$ we obtain estimate values for $\alpha$ and $\beta$ that we
respectively denote by $\alpha^{\ast}$ and $\beta^{\ast}.$ Then, we use the
first equation of (\ref{c-d}) to get the corresponding estimate $c^{\ast}$ of
$c.$

\noindent\textbf{Step 3:} Using formulas (\ref{newestim}) and (\ref{var}), we
compute $\mu^{\ast}=\widehat{\mu}\left(  k^{\ast}\right)  $ and $\sigma\left(
\alpha^{\ast},\beta^{\ast}\right)  $ respectively.

\begin{figure}
[h]
\begin{center}
\includegraphics[
height=4.209in,
width=4.7625in
]%
{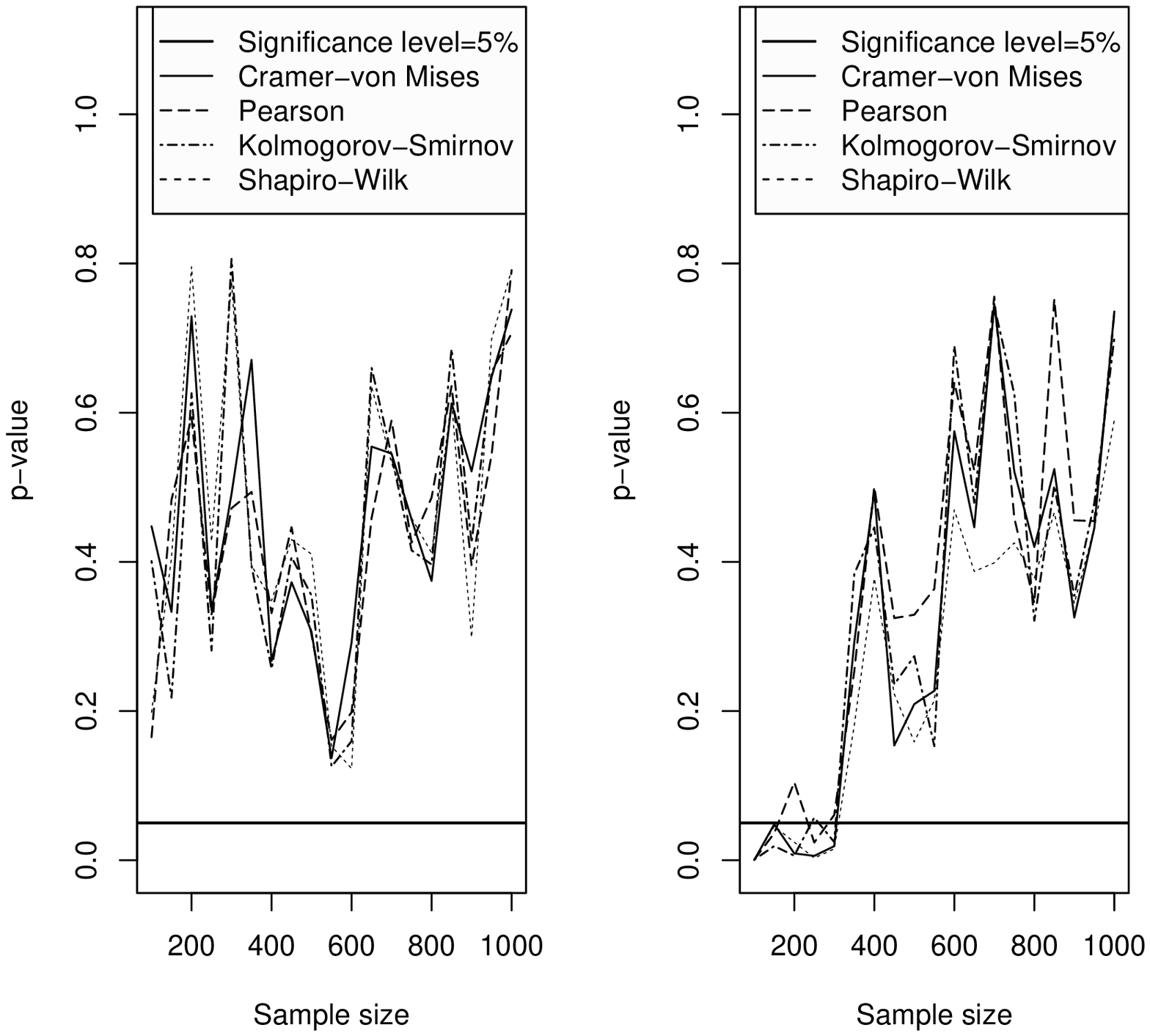}%
\caption{Empirical p-values of normality tests for the new estimator (left
panel) and Peng's estimator (right panel) based on $200$ samples of a
Fr\'{e}chet population with shape parameter $\alpha=1.5.$}%
\label{FIG15}%
\end{center}
\end{figure}

\begin{figure}
[ptb]
\begin{center}
\includegraphics[
height=4.7383in,
width=4.7504in
]%
{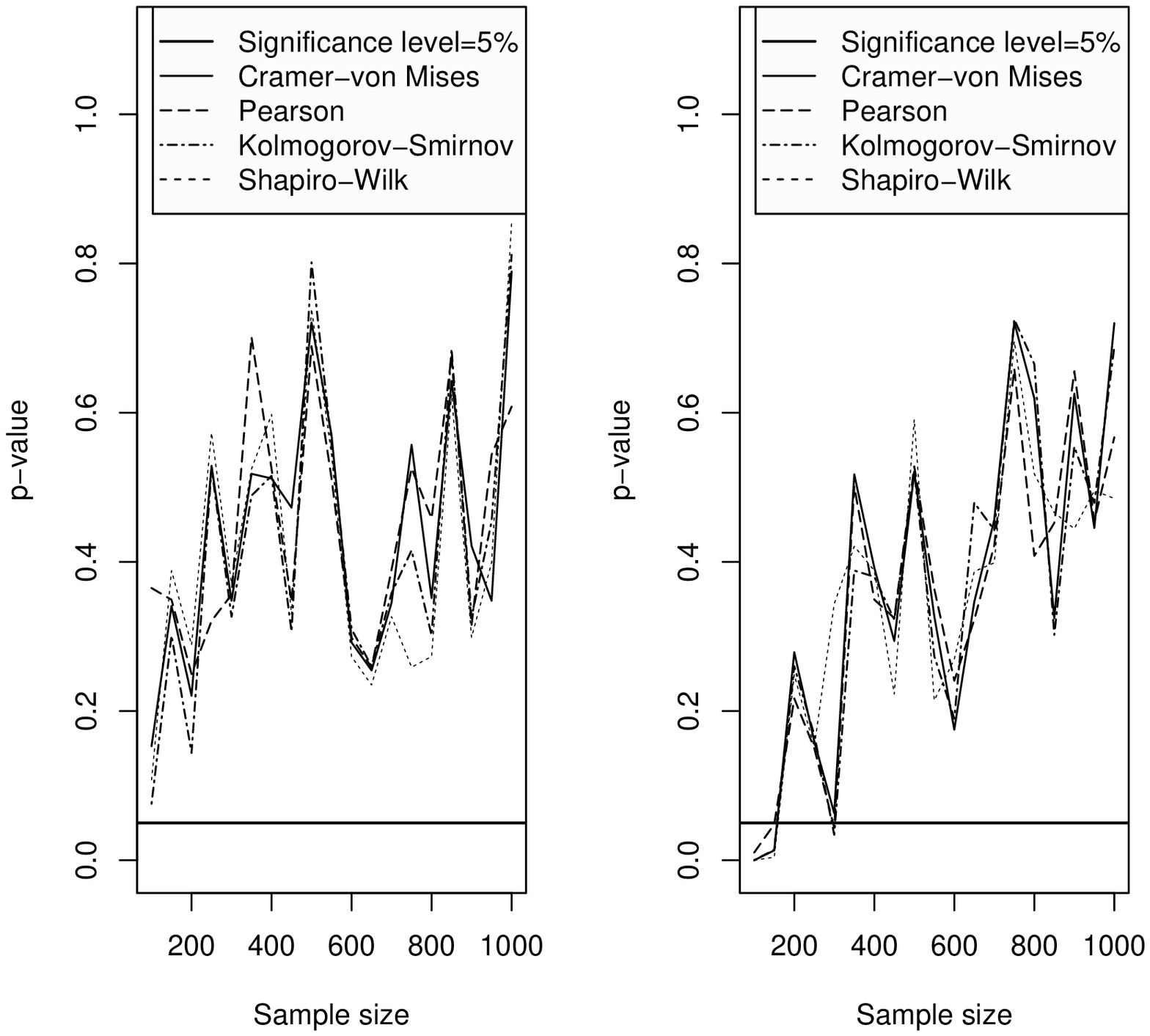}%
\caption{Empirical p-values of normality tests for the new estimator (left
panel) and Peng's estimator (right panel) based on $200$ samples from a
Fr\'{e}chet population with shape parameter $\alpha=1.7.$}%
\label{FIG17}%
\end{center}
\end{figure}

\begin{figure}
[ptb]
\begin{center}
\includegraphics[
height=4.7383in,
width=4.7504in
]%
{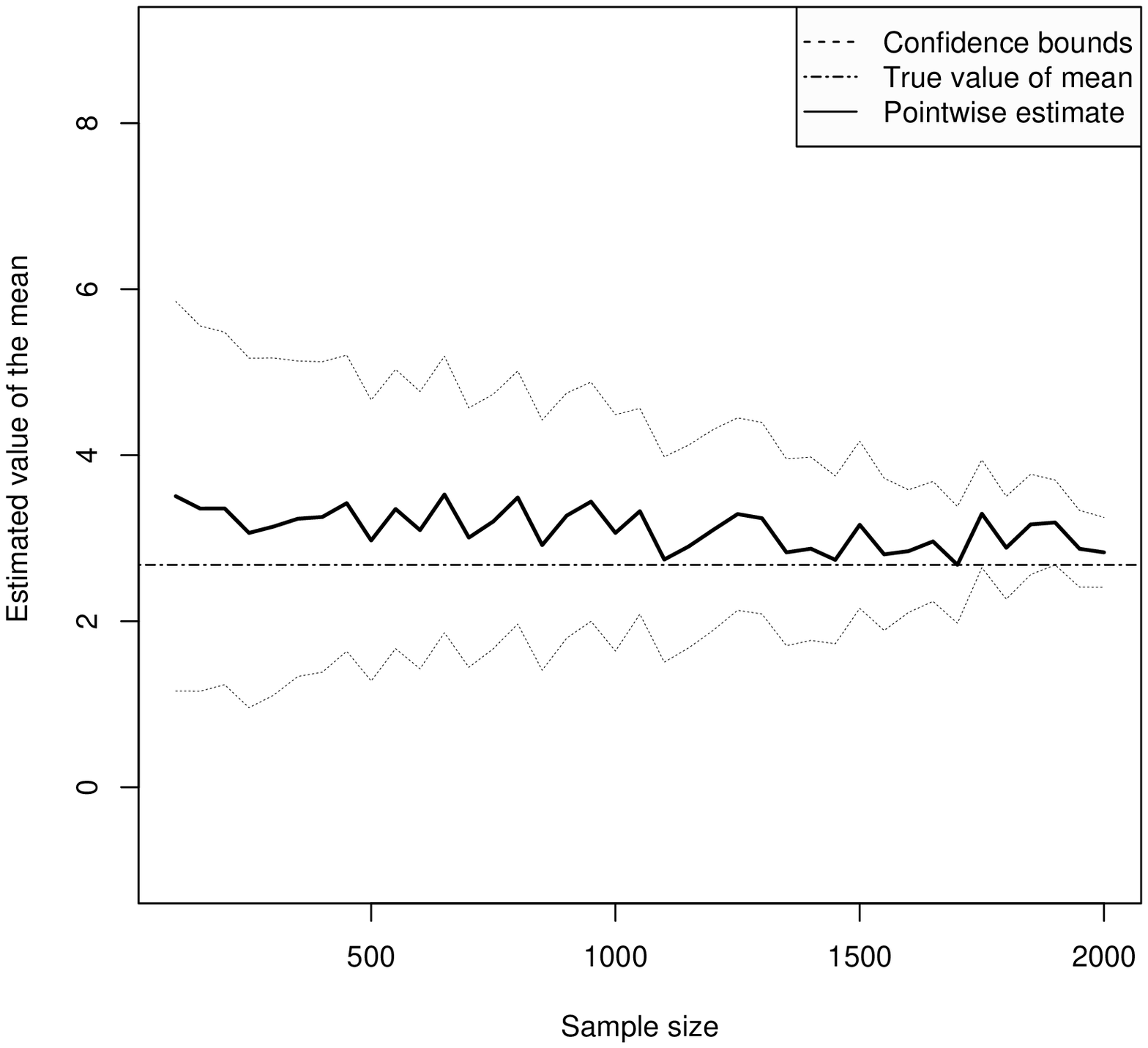}%
\caption{Confidence intervals for the mean $\mu$ based on $200$ samples of
size $n,$ ranging from $100$ to $2000,$ from a Fr\'{e}chet population with
shape parameter $\alpha=1.5.$ The horizontal line represents the true value
$\mu=2.678.$}%
\label{FIG-CB}%
\end{center}
\end{figure}

\noindent\textbf{Step 4:}\ Finally, Corollary \ref{corollary} yields the
$\left(  1-\zeta/2\right)  100\%$ confidence interval for $\mu:$
\[
\mu^{\ast}\pm z_{\zeta}\frac{\sqrt{k^{\ast}/n}\sigma\left(  \alpha^{\ast
},\beta^{\ast}\right)  \left(  nc^{\ast}/k^{\ast}\right)  ^{1/\alpha^{\ast}}%
}{\sqrt{n}}%
\]
Our simulation study, which is based on $200$ samples of various sizes from
the Fr\'{e}chet distribution (\ref{Frechet}) with two distinct tail index
values $\alpha=1.5$ and $1.7,$ consists of three parts. First, we compare, in
terms of bias and root of the mean squared error (RMSE), the performances of
the new estimator $\widehat{\mu}_{n}$ and Peng's estimator $\widehat{\mu}%
_{n}^{P}.$ The results of this part are summarized in Tables \ref{Tab1A} and
\ref{Tab1B}. Second, we check the asymptotic normality of both estimators via
four of the most popular goodness-of-fit tests at the $5\%$ significance
level: Cram\'{e}r-von Mises (CvM), Kolmogrov-Smirnov (KS), Shapiro-Wilk (SW)
and Pearson (P). The results of this part are summarized in Tables \ref{Tab3}
and \ref{Tab4} and illustrated by Figures \ref{FIG15} and \ref{FIG17}.
Finally, we investigate the accuracy of the confidence intervals, built from
the new estimator $\widehat{\mu}_{n},$ by computing their lengths and coverage
probabilities (denoted by `covpr'). The results of this part are summarized in
Table \ref{Tab2} (where `lcb' and `ucb' respectively stand for the lower and
upper confidence bounds) and illustrated by Figure \ref{FIG-CB}.%

\begin{table}[h] \centering
\begin{tabular}
[c]{lrrrrrrrr}
& \multicolumn{4}{c}{$\widehat{\mu}_{n}$} & \multicolumn{4}{c}{$\widehat{\mu
}_{n}^{P}$}\\\hline
Sample size & ${\small 500}$ & ${\small 1000}$ & ${\small 2000}$ &
${\small 3000}$ & ${\small 500}$ & ${\small 1000}$ & ${\small 2000}$ &
${\small 3000}$\\\hline\hline
Estimated value & ${\small 3.089}$ & ${\small 2.999}$ & ${\small 2.912}$ &
${\small 2.467}$ & ${\small 3.367}$ & ${\small 3.119}$ & ${\small 3.119}$ &
${\small 2.289}$\\
Bias & ${\small 0.411}$ & ${\small 0.321}$ & ${\small 0.234}$ &
${\small 0.204}$ & ${\small 0.689}$ & ${\small 0.441}$ & ${\small 0.441}$ &
${\small 0.389}$\\
RMSE & ${\small 0.400}$ & ${\small 0.286}$ & ${\small 0.125}$ &
${\small 0.108}$ & ${\small 0.674}$ & ${\small 0.268}$ & ${\small 0.268}$ &
${\small 0.198}$\\\hline\hline
&  &  &  &  &  &  &  &
\end{tabular}
\caption{Point estimation of the mean based on 200 samples from the
Fr\'{e}chet population with shape parameter $\alpha=1.5$. The true value of the mean is 2.678.}\label{Tab1A}%
\end{table}%
%

\begin{table}[h] \centering
\begin{tabular}
[c]{lrrrrrrrr}
& \multicolumn{4}{c}{$\widehat{\mu}_{n}$} & \multicolumn{4}{c}{$\widehat{\mu
}_{n}^{P}$}\\\hline
Sample size & ${\small 500}$ & ${\small 1000}$ & ${\small 2000}$ &
${\small 3000}$ & ${\small 500}$ & ${\small 1000}$ & ${\small 2000}$ &
${\small 3000}$\\\hline\hline
Estimated value & ${\small 2.536}$ & ${\small 2.440}$ & ${\small 2.354}$ &
${\small 2.254}$ & ${\small 2.765}$ & ${\small 2.731}$ & ${\small 2.566}$ &
${\small 2.460}$\\
Bias & ${\small 0.383}$ & ${\small 0.287}$ & ${\small 0.201}$ &
${\small 0.101}$ & ${\small 0.612}$ & ${\small 0.578}$ & ${\small 0.413}$ &
${\small 0.307}$\\
RMSE & ${\small 0.340}$ & ${\small 0.211}$ & ${\small 0.114}$ &
${\small 0.089}$ & ${\small 0.659}$ & ${\small 0.420}$ & ${\small 0.250}$ &
${\small 0.144}$\\\hline\hline
&  &  &  &  &  &  &  &
\end{tabular}
\caption{Point estimation of the mean based on 200 samples from the
Fr\'{e}chet population with shape parameter $\alpha=1.7$. The true value of the mean is 2.153.}\label{Tab1B}%
\end{table}%
%

\begin{table}[h] \centering
\begin{tabular}
[c]{ccccc}
&  &  &  & \\\hline
{\small Sample size} & {\small CvM} & {\small KS} & {\small SW} &
\multicolumn{1}{c|}{{\small P}}\\\hline\hline
${\small 100}$ & ${\small 0.447}$ & ${\small 0.401}$ & ${\small 0.198}$ &
\multicolumn{1}{c|}{${\small 0.165}$}\\
${\small 200}$ & ${\small 0.729}$ & ${\small 0.626}$ & ${\small 0.795}$ &
\multicolumn{1}{c|}{${\small 0.549}$}\\
${\small 400}$ & ${\small 0.267}$ & ${\small 0.256}$ & ${\small 0.347}$ &
\multicolumn{1}{c|}{${\small 0.331}$}\\
${\small 500}$ & ${\small 0.306}$ & ${\small 0.354}$ & ${\small 0.410}$ &
\multicolumn{1}{c|}{${\small 0.302}$}\\
${\small 800}$ & ${\small 0.374}$ & ${\small 0.396}$ & ${\small 0.412}$ &
\multicolumn{1}{c|}{${\small 0.486}$}\\
${\small 1000}$ & ${\small 0.738}$ & ${\small 0.706}$ & ${\small 0.691}$ &
\multicolumn{1}{c|}{${\small 0.722}$}\\\hline\hline
&  &  &  &
\end{tabular}%
\begin{tabular}
[c]{ccccc}
&  &  &  & \\\hline
\multicolumn{1}{|c}{{\small Sample size}} & {\small CvM} & {\small KS} &
{\small SW} & {\small P}\\\hline\hline
\multicolumn{1}{|c}{${\small 100}$} & ${\small 0.001}$ & ${\small 0.000}$ &
${\small 0.000}$ & ${\small 0.002}$\\
\multicolumn{1}{|c}{${\small 200}$} & ${\small 0.009}$ & ${\small 0.006}$ &
${\small 0.024}$ & ${\small 0.104}$\\
\multicolumn{1}{|c}{${\small 400}$} & ${\small 0.495}$ & ${\small 0.446}$ &
${\small 0.377}$ & ${\small 0.500}$\\
\multicolumn{1}{|c}{${\small 500}$} & ${\small 0.209}$ & ${\small 0.273}$ &
${\small 0.158}$ & ${\small 0.329}$\\
\multicolumn{1}{|c}{${\small 800}$} & ${\small 0.419}$ & ${\small 0.321}$ &
${\small 0.378}$ & ${\small 0.344}$\\
\multicolumn{1}{|c}{${\small 1000}$} & ${\small 0.724}$ & ${\small 0.711}$ &
${\small 0.590}$ & ${\small 0.733}$\\\hline\hline
&  &  &  &
\end{tabular}
\caption{Empirical p-values of normality tests for the new estimator (left panel) and Peng's estimator (right panel)
based on 200 samples from a Fr\'{e}chet population with shape parameter $\alpha=1.5$.}\label{Tab3}%
\end{table}%
%

\begin{table}[h] \centering
\begin{tabular}
[c]{ccccc}
&  &  &  & \\\hline
{\small Sample size} & {\small CvM} & {\small KS} & {\small SW} &
\multicolumn{1}{c|}{{\small P}}\\\hline\hline
${\small 100}$ & ${\small 0.153}$ & ${\small 0.076}$ & ${\small 0.107}$ &
\multicolumn{1}{c|}{${\small 0.364}$}\\
${\small 200}$ & ${\small 0.220}$ & ${\small 0.143}$ & ${\small 0.288}$ &
\multicolumn{1}{c|}{${\small 0.249}$}\\
${\small 400}$ & ${\small 0.511}$ & ${\small 0.515}$ & ${\small 0.397}$ &
\multicolumn{1}{c|}{${\small 0.524}$}\\
${\small 500}$ & ${\small 0.713}$ & ${\small 0.781}$ & ${\small 0.624}$ &
${\small 0.635}$\\
${\small 800}$ & ${\small 0.362}$ & ${\small 0.311}$ & ${\small 0.261}$ &
\multicolumn{1}{c|}{${\small 0.458}$}\\
${\small 1000}$ & ${\small 0.778}$ & ${\small 0.783}$ & ${\small 0.645}$ &
\multicolumn{1}{c|}{${\small 0.601}$}\\\hline\hline
&  &  &  &
\end{tabular}%
\begin{tabular}
[c]{ccccc}
&  &  &  & \\\hline
\multicolumn{1}{|c}{{\small Sample size}} & {\small CvM} & {\small KS} &
{\small SW} & {\small P}\\\hline\hline
\multicolumn{1}{|c}{${\small 100}$} & ${\small 0.013}$ & ${\small 0.014}$ &
${\small 0.004}$ & ${\small 0.010}$\\
\multicolumn{1}{|c}{${\small 200}$} & ${\small 0.278}$ & ${\small 0.260}$ &
${\small 0.248}$ & ${\small 0.216}$\\
\multicolumn{1}{|c}{${\small 400}$} & ${\small 0.392}$ & ${\small 0.380}$ &
${\small 0.298}$ & ${\small 0.349}$\\
\multicolumn{1}{|c}{${\small 500}$} & ${\small 0.520}$ & ${\small 0.492}$ &
${\small 0.480}$ & ${\small 0.528}$\\
\multicolumn{1}{|c}{${\small 800}$} & ${\small 0.619}$ & ${\small 0.665}$ &
${\small 0.408}$ & ${\small 0.518}$\\
\multicolumn{1}{|c}{${\small 1000}$} & ${\small 0.720}$ & ${\small 0.688}$ &
${\small 0.485}$ & ${\small 0.567}$\\\hline\hline
&  &  &  &
\end{tabular}
\caption{Empirical p-values of normality tests for the new estimator (left panel) and Peng's estimator (right panel)
based on 200 samples from a Fr\'{e}chet population with shape parameter $\alpha=1.7$.}\label{Tab4}%
\end{table}%
$\medskip$%

\begin{table}[h] \centering
\begin{tabular}
[c]{ccccccclllll}\hline
${\small n}$ & {\small lcb} & $\widehat{\mu}$ & {\small ucb} & {\small covpr}
& {\small length} & \multicolumn{1}{|c}{${\small n}$} & {\small lcb} &
$\widehat{\mu}$ & {\small ucb} & {\small covpr} & {\small length}%
\\\hline\hline
${\small 100}$ & \multicolumn{1}{r}{${\small -0.239}$} &
\multicolumn{1}{r}{${\small 2.992}$} & \multicolumn{1}{r}{${\small 3.793}$} &
\multicolumn{1}{r}{${\small 0.666}$} & \multicolumn{1}{r|}{${\small 4.032}$} &
\multicolumn{1}{|c}{${\small 100}$} & \multicolumn{1}{r}{${\small -0.524}$} &
\multicolumn{1}{r}{${\small 2.823}$} & \multicolumn{1}{r}{${\small 3.281}$} &
\multicolumn{1}{r}{${\small 0.566}$} & \multicolumn{1}{r}{${\small 3.805}$}\\
${\small 200}$ & \multicolumn{1}{r}{${\small 0.971}$} &
\multicolumn{1}{r}{${\small 3.071}$} & \multicolumn{1}{r}{${\small 3.623}$} &
\multicolumn{1}{r}{${\small 0.445}$} & \multicolumn{1}{r|}{${\small 2.652}$} &
\multicolumn{1}{|c}{${\small 200}$} & \multicolumn{1}{r}{${\small -0.283}$} &
\multicolumn{1}{r}{${\small 2.794}$} & \multicolumn{1}{r}{${\small 3.257}$} &
\multicolumn{1}{r}{${\small 0.681}$} & \multicolumn{1}{r}{${\small 3.540}$}\\
${\small 400}$ & \multicolumn{1}{r}{${\small 1.485}$} &
\multicolumn{1}{r}{${\small 3.061}$} & \multicolumn{1}{r}{${\small 3.602}$} &
\multicolumn{1}{r}{${\small 0.603}$} & \multicolumn{1}{r|}{${\small 2.117}$} &
\multicolumn{1}{|c}{${\small 400}$} & \multicolumn{1}{r}{${\small 0.760}$} &
\multicolumn{1}{r}{${\small 2.766}$} & \multicolumn{1}{r}{${\small 3.197}$} &
\multicolumn{1}{r}{${\small 0.700}$} & \multicolumn{1}{r}{${\small 2.437}$}\\
${\small 500}$ & \multicolumn{1}{r}{${\small 1.497}$} &
\multicolumn{1}{r}{${\small 3.026}$} & \multicolumn{1}{r}{${\small 3.565}$} &
\multicolumn{1}{r}{${\small 0.666}$} & \multicolumn{1}{r|}{${\small 2.068}$} &
\multicolumn{1}{|c}{${\small 500}$} & \multicolumn{1}{r}{${\small 0.770}$} &
\multicolumn{1}{r}{${\small 2.544}$} & \multicolumn{1}{r}{${\small 3.154}$} &
\multicolumn{1}{r}{${\small 0.733}$} & \multicolumn{1}{r}{${\small 2.384}$}\\
${\small 800}$ & \multicolumn{1}{r}{${\small 1.708}$} &
\multicolumn{1}{r}{${\small 2.943}$} & \multicolumn{1}{r}{${\small 3.403}$} &
\multicolumn{1}{r}{${\small 0.785}$} & \multicolumn{1}{r|}{${\small 1.695}$} &
\multicolumn{1}{|c}{${\small 800}$} & \multicolumn{1}{r}{${\small 0.712}$} &
\multicolumn{1}{r}{${\small 2.471}$} & \multicolumn{1}{r}{${\small 3.013}$} &
\multicolumn{1}{r}{${\small 0.800}$} & \multicolumn{1}{r}{${\small 2.301}$}\\
${\small 1000}$ & \multicolumn{1}{r}{${\small 1.922}$} &
\multicolumn{1}{r}{${\small 2.883}$} & \multicolumn{1}{r}{${\small 3.302}$} &
\multicolumn{1}{r}{${\small 0.795}$} & \multicolumn{1}{r|}{${\small 1.380}$} &
\multicolumn{1}{|c}{${\small 1000}$} & \multicolumn{1}{r}{${\small 1.061}$} &
\multicolumn{1}{r}{${\small 2.451}$} & \multicolumn{1}{r}{${\small 2.492}$} &
\multicolumn{1}{r}{${\small 0.833}$} & \multicolumn{1}{r}{${\small 1.431}$%
}\\\hline\hline
&  &  &  &  &  &  &  &  &  &  &
\end{tabular}
\caption{Accuracy of $95\%$ confidence intervals for the new estimator based
on 200 samples from Fr\'{e}chet populations with shape parameters $\alpha=1.5$ (left panel) and
$\alpha=1.7$  (right panel).}.\label{Tab2}%
\end{table}%

%

%

%

\noindent Tables \ref{Tab1A} and \ref{Tab1B} show that, regardless of the
sample size, our new estimator performs better than Peng's one as far as the
bias and RMSE are concerned. Moreover, from the left panels of Figure
\ref{FIG17} and Table \ref{Tab4} (corresponding to the case of the lighter
tail $\alpha=1.7)$ we see that the normality of $\widehat{\mu}_{n}$ cannot be
rejected by any of the tests when the sample size exceeds $100,$ while the
right panels of Figure \ref{FIG17} and Table \ref{Tab4} show that the
normality of $\widehat{\mu}_{n}^{P}$ is rejected for sample sizes ranging
between $100$ and $200.$ In the case of the heavier tail $\alpha=1.5,$ the
right panels of Figure \ref{FIG15} and Table \ref{Tab3} show that the sample
size needs to be larger then $400$ for the estimator $\widehat{\mu}_{n}^{P}$
to pass the normality tests, while the left panels of Figure \ref{FIG15} and
Table \ref{Tab3} indicate that the normality of $\widehat{\mu}_{n}$ is
accepted even for sample sizes smaller than $200.$

\section{\textbf{Proofs\label{secProof}}}

\subsection{Proof \textbf{of Theorem} \textbf{\ref{Theorem1}.}}

First recall that $\lambda=\beta/\alpha>1.\ $Then, from expansion $\left(
4.1\right)  $ in \cite{LiPeYa10}, we have, as $n\rightarrow\infty$%
\begin{equation}
\widehat{\alpha}-\alpha=\alpha\left\{  \eta_{1}\left(  S_{1}-1\right)
+\eta_{2}\left(  S_{2}-\lambda^{-1}\right)  +\eta_{3}\left(  S_{3}%
-\lambda^{-2}\right)  \right\}  +o_{p}\left(  k^{-1/2}\right)  , \label{main}%
\end{equation}
where%
\begin{equation}
S_{1}:=\dfrac{1}{k}%
{\displaystyle\sum\limits_{i=1}^{k}}
\log\dfrac{Y_{n-i+1,n}}{Y_{n-k,n}},\text{ }S_{2}:=\dfrac{1}{k}%
{\displaystyle\sum\limits_{i=1}^{k}}
\left(  \dfrac{Y_{n-i+1,n}}{Y_{n-k,n}}\right)  ^{1-\lambda}, \label{P0}%
\end{equation}
and%
\begin{equation}
S_{3}:=\dfrac{1}{k}\sum\limits_{i=1}^{k}\left(  \dfrac{Y_{n-i+1,n}}{Y_{n-k,n}%
}\right)  ^{1-\lambda}\log\dfrac{Y_{n-i+1,n}}{Y_{n-k},_{n}}, \label{P}%
\end{equation}
with $Y_{1,n}\leq...\leq Y_{n,n}$ being the order statistics pertaining to a
sample $Y_{1},...,Y_{n}$ of i.i.d. r.v.'s, defined on the same probability
space as the $X_{i}^{\prime}s,$ with cdf
\begin{equation}
G\left(  y\right)  =1-y^{-1},\text{ for }y>1. \label{G}%
\end{equation}
\cite{CsCsHM86} have constructed a probability space $\left(  \Omega
,\emph{A,}\mathbf{P}\right)  $ carrying an infinite sequence $U_{1},U_{2},...$
of independent $\left(  0,1\right)  -$uniform r.v.'s and a sequence of
Brownian bridges $\left\{  B_{n}\left(  s\right)  ,\text{ }0\leq
s\leq1\right\}  ,$ $n=1,2,...,$ having, amongst others, the property stated in
Lemma \ref{Lemma1}. Let $U_{1,n}\leq...\leq U_{n,n}$ denote the order
statistics pertaining to $U_{1},...,U_{n}$ and define the empirical quantile
function $V_{n}\left(  s\right)  $ as%
\[
V_{n}\left(  s\right)  =U_{i,n}\quad\text{for}\ \left(  i-1\right)  /n<s\leq
i/n,\text{ }i=1,...,n,\text{ and }V_{n}\left(  0\right)  =U_{1,n}.
\]

\begin{lemma}
\label{Lemma1}On the probability space of \cite{CsCsHM86}, for every
$0\leq\tau<1/2,$ we have, as $n\rightarrow\infty$%
\begin{equation}
\sup_{1/n\leq s\leq1-1/n}\frac{\left\vert \sqrt{n}\left(  s-V_{n}\left(
s\right)  \right)  -B_{n}\left(  s\right)  \right\vert }{\left(  s\left(
1-s\right)  \right)  ^{1/2-\tau}}=O_{p}\left(  n^{-\tau}\right)  .
\label{approxi}%
\end{equation}

\end{lemma}

\noindent\textbf{Proof.} See the proof of Theorem 2.1 in \cite{CsCsHM86}%
.\hfill$\square$

\noindent Without loss of generality, we assume that%
\[
Y_{i}=G^{-1}\left(  U_{i}\right)  =\left(  1-U_{i}\right)  ^{-1},\text{
}i=1,...,n,
\]
and%
\[
Y_{i,n}=G^{-1}\left(  U_{i,n}\right)  =\left(  1-U_{i,n}\right)  ^{-1},\text{
}i=1,...,n,
\]
where $G^{-1}$ denotes the quantile function pertaining to cdf $G$ given by
formula (\ref{G}).$\ $Then, this allows us to write%
\[
Y_{n-i+1,n}=\left(  1-V_{n}\left(  1-s\right)  \right)  ^{-1},\text{ for
}\dfrac{i-1}{n}<s\leq\dfrac{i}{n},\text{ }i=1,...,n.
\]
Making use of the previous representation of the order statistics
$Y_{n-i+1,n}$, we may rewrite the three statistics in (\ref{P0}) and (\ref{P})
into%
\begin{align*}
S_{1}  &  =\frac{n}{k}\int_{0}^{k/n}\log\left(  \frac{1-V_{n}\left(
1-s\right)  }{1-U_{n-k,n}}\right)  ^{-1}ds,\\
S_{2}  &  =\frac{n}{k}\int_{0}^{k/n}\left(  \frac{1-V_{n}\left(  1-s\right)
}{1-U_{n-k,n}}\right)  ^{-1+\lambda}ds,
\end{align*}
and%
\[
S_{3}=\frac{n}{k}\int_{0}^{k/n}\left(  \frac{1-V_{n}\left(  1-s\right)
}{1-U_{n-k,n}}\right)  ^{-1+\lambda}\log\left(  \frac{1-V_{n}\left(
1-s\right)  }{1-U_{n-k,n}}\right)  ^{-1}ds.
\]
Next, we show that, as $n\rightarrow\infty$%
\begin{align*}
\sqrt{k}\left(  S_{1}-1\right)   &  =W_{1n}+o_{p}\left(  1\right)  ,\\
\sqrt{k}\left(  S_{2}-\lambda^{-1}\right)   &  =W_{2n}+o_{p}\left(  1\right)
,
\end{align*}
and%
\[
\sqrt{k}\left(  S_{3}-\lambda^{-2}\right)  =W_{3n}+o_{p}\left(  1\right)  ,
\]
where $W_{1n},W_{2n}$ and $W_{3n}$ are the Gaussian r.v.'s defined in Theorem
\ref{Theorem1}.$\ $We will only consider the asymptotic distribution of
$S_{3}.$ The proofs for $S_{1}$ and $S_{2}$ use similar arguments. By letting
$f\left(  x\right)  =x^{\lambda-1}\log x,$ the statistic $S_{3}$ becomes%
\[
S_{3}=-\left(  n/k\right)
{\displaystyle\int_{0}^{k/n}}
f\left(  \dfrac{1-V_{n}\left(  1-s\right)  }{1-U_{n-k,n}}\right)  ds.
\]
An application of standard calculus gives $\int_{0}^{1}f\left(  s\right)
ds=-\lambda^{-2}.$ Therefore%
\[
S_{3}-\lambda^{-2}=-\left(  n/k\right)
{\displaystyle\int_{0}^{k/n}}
\left[  f\left(  \dfrac{1-V_{n}\left(  1-s\right)  }{1-U_{n-k,n}}\right)
-f\left(  \frac{s}{k/n}\right)  \right]  ds.
\]
Let us follow similar techniques as those used in the proof of Lemma 9 in
\cite{CsDeMa85}. We divide the integral above in two parts, then we study the
asymptotic behavior of each integral. Observe that%
\begin{align*}
S_{3}-\lambda^{-2}  &  =-\left(  n/k\right)
{\displaystyle\int_{0}^{1/n}}
\left[  f\left(  \dfrac{1-V_{n}\left(  1-s\right)  }{1-U_{n-k,n}}\right)
-f\left(  \frac{s}{k/n}\right)  \right]  ds\\
&  \ \ \ \ \ \ \ -\left(  n/k\right)
{\displaystyle\int_{1/n}^{k/n}}
\left[  f\left(  \dfrac{1-V_{n}\left(  1-s\right)  }{1-U_{n-k,n}}\right)
-f\left(  \frac{s}{k/n}\right)  \right]  ds\\
&
\begin{array}
[c]{c}%
=:
\end{array}
-\Delta_{n}-\Omega_{n}.
\end{align*}
Next, we show that $\sqrt{k}\Delta_{n}$ converges to $0$ in probability.
Indeed, we have $1-V_{n}\left(  1-s\right)  =1-U_{n,n},$ for $0<s\leq1/n,$ it
follows that%
\begin{align*}
\Delta_{n}  &  =\left(  n/k\right)
{\displaystyle\int_{0}^{1/n}}
\left[  f\left(  \dfrac{1-U_{n,n}}{1-U_{n-k,n}}\right)  -f\left(  \frac
{s}{k/n}\right)  \right]  ds\\
&  =k^{-1}f\left(  \dfrac{1-U_{n,n}}{1-U_{n-k,n}}\right)  -%
{\displaystyle\int_{0}^{1/k}}
f\left(  s\right)  ds.
\end{align*}
An elementary calculation gives $%
{\displaystyle\int_{0}^{1/k}}
f\left(  s\right)  ds=\lambda^{-1}k^{-\lambda}\left(  \log k^{-1}-\lambda
^{-1}\right)  ,$ and from Lemma 2.2.3 of page 41 in \cite{deHFe06}, we have
$n\left(  1-U_{n,n-k}\right)  /k\overset{P}{\rightarrow}1,$ as $n\rightarrow
\infty,$ therefore%
\[
\Delta_{n}=\left\{  1+o_{p}\left(  1\right)  \right\}  k^{-\lambda}\log
k^{-1}-\lambda^{-1}k^{-\lambda}\left(  \log k^{-1}-\lambda^{-1}\right)  .
\]
Since $\lambda>1$ and $k\rightarrow\infty,$ then $k^{-\lambda+1/2}%
\rightarrow0$ and $k^{-\lambda+1/2}\log k^{-1}\rightarrow0,$ it follows that
$\sqrt{k}\Delta_{n}\overset{P}{\rightarrow}0$ as $n\rightarrow\infty.$
Consider now the second term $\Omega_{n}$ which may be rewritten into%
\begin{align*}
\Omega_{n}  &  =\left(  n/k\right)
{\displaystyle\int_{1/n}^{k/n}}
\left[  f\left(  \dfrac{1-V_{n}\left(  1-s\right)  }{1-U_{n-k,n}}\right)
-f\left(  \dfrac{s}{1-U_{n-k,n}}\right)  \right]  ds\\
&  \ \ \ \ \ \ \ \ \ \ \ \ \ \ \ \ \ +\left(  n/k\right)
{\displaystyle\int_{1/n}^{k/n}}
\left[  f\left(  \dfrac{s}{1-U_{n-k,n}}\right)  -f\left(  \dfrac{s}%
{k/n}\right)  \right]  ds\\
&
\begin{array}
[c]{c}%
=:
\end{array}
\Omega_{n1}+\Omega_{n2}.
\end{align*}
Making use of Taylor's expansion of $f,$ we get%
\[
f\left(  \dfrac{1-V_{n}\left(  1-s\right)  }{1-U_{n-k,n}}\right)  -f\left(
\dfrac{s}{1-U_{n-k,n}}\right)  =f^{\prime}\left(  \frac{\varphi_{n}\left(
s\right)  }{1-U_{n-k,n}}\right)  \left(  \dfrac{1-V_{n}\left(  1-s\right)
}{1-U_{n-k,n}}-\dfrac{s}{1-U_{n-k,n}}\right)  ,
\]
and%
\[
f\left(  \dfrac{s}{1-U_{n-k,n}}\right)  -f\left(  \dfrac{s}{k/n}\right)
=f^{\prime}\left(  s\psi_{n}\right)  \left(  \dfrac{s}{1-U_{n-k,n}}-\dfrac
{s}{k/n}\right)  ,
\]
where%
\begin{equation}
\min\left\{  1-V_{n}\left(  1-s\right)  ,s\right\}  <\varphi_{n}\left(
s\right)  <\max\left\{  1-V_{n}\left(  1-s\right)  ,s\right\}  \label{phi}%
\end{equation}
and%
\begin{equation}
\min\left\{  \dfrac{1}{1-U_{n-k,n}},\dfrac{1}{k/n}\right\}  <\psi_{n}%
<\max\left\{  \dfrac{1}{1-U_{n-k,n}},\dfrac{1}{k/n}\right\}  . \label{psi}%
\end{equation}
Observe now that $\Omega_{n1}$ and $\Omega_{n2}$ may be rewritten into
\[
\Omega_{n1}=\left(  n/k\right)
{\displaystyle\int_{1/n}^{k/n}}
f^{\prime}\left(  \dfrac{s}{1-U_{n-k,n}}\right)  \left[  \dfrac{1-V_{n}\left(
1-s\right)  }{1-U_{n-k,n}}-\dfrac{s}{1-U_{n-k,n}}\right]  ds+\Omega_{n1}%
^{\ast},
\]
and%
\[
\Omega_{n2}=\left\{  1+o_{p}\left(  1\right)  \right\}  \left(  n/k\right)
{\displaystyle\int_{1/n}^{k/n}}
f^{\prime}\left(  \dfrac{s}{k/n}\right)  \left[  \dfrac{s}{1-U_{n-k,n}}%
-\dfrac{s}{k/n}\right]  ds+\Omega_{n2}^{\ast},
\]
where
\begin{align*}
&  \Omega_{n1}^{\ast}%
\begin{array}
[c]{c}%
:=
\end{array}
\left(  n/k\right)
{\displaystyle\int_{1/n}^{k/n}}
\left[  f^{\prime}\left(  \dfrac{\varphi_{n}\left(  s\right)  }{1-U_{n-k,n}%
}\right)  -f^{\prime}\left(  \dfrac{s}{1-U_{n-k,n}}\right)  \right] \\
&
\ \ \ \ \ \ \ \ \ \ \ \ \ \ \ \ \ \ \ \ \ \ \ \ \ \ \ \ \ \ \ \ \ \ \ \ \ \ \ \ \ \times
\left[  \dfrac{1-V_{n}\left(  1-s\right)  }{1-U_{n-k,n}}-\dfrac{s}%
{1-U_{n-k,n}}\right]  ds,
\end{align*}
and%
\begin{align*}
&  \Omega_{n2}^{\ast}%
\begin{array}
[c]{c}%
:=
\end{array}
\left\{  1+o_{p}\left(  1\right)  \right\}  \left(  n/k\right)
{\displaystyle\int_{1/n}^{k/n}}
\left[  f^{\prime}\left(  \dfrac{s\psi_{n}}{k/n}\right)  -f^{\prime}\left(
\dfrac{s}{k/n}\right)  \right] \\
&
\ \ \ \ \ \ \ \ \ \ \ \ \ \ \ \ \ \ \ \ \ \ \ \ \ \ \ \ \ \ \ \ \ \ \ \ \ \ \ \ \ \ \ \ \ \ \ \ \times
\left[  \dfrac{s}{1-U_{n-k,n}}-\dfrac{s}{k/n}\right]  ds.
\end{align*}
From Lemma \ref{Lemma3} (see the Appendix), both $\sqrt{k}\Omega_{n1}^{\ast}$
and $\sqrt{k}\Omega_{n2}^{\ast}$ converge to $0$ in probability. Since
$n\left(  1-U_{n-k,n}\right)  /k\overset{P}{\rightarrow}1,$ then%
\[
\Omega_{n1}=\left\{  1+o_{p}\left(  1\right)  \right\}
{\displaystyle\int_{1/n}^{k/n}}
f^{\prime}\left(  \dfrac{s}{k/n}\right)  \left[  1-s-V_{n}\left(  1-s\right)
\right]  ds+o_{p}\left(  k^{-1/2}\right)  ,
\]
and%
\[
\Omega_{n2}=-\left\{  1+o_{p}\left(  1\right)  \right\}  \dfrac{k/n-\left(
1-U_{n-k,n}\right)  }{\left(  k/n\right)  ^{2}}%
{\displaystyle\int_{1/n}^{k/n}}
\left(  \dfrac{s}{k/n}\right)  f^{\prime}\left(  \dfrac{s}{k/n}\right)
ds+o_{p}\left(  k^{-1/2}\right)  .
\]
The derivative of function $f$ equals $f^{\prime}\left(  x\right)
=(\lambda-1)x^{\lambda-2}\log x+x^{\lambda-2},$ then
\begin{align*}
\Omega_{n1}  &  =\left(  \lambda-1\right)  \left(  n/k\right)
{\displaystyle\int_{1/k}^{1}}
t^{\lambda-2}\left(  \log t\right)  \left[  1-V_{n}\left(  1-kt/n\right)
-kt/n\right]  dt\\
&  \ \ \ \ \ \ \ \ \ \ \ \ \ \ \ \ \ \ \ \ \ \ \ \ \ \ +\left(  n/k\right)
{\displaystyle\int_{1/k}^{1}}
t^{\lambda-2}\left[  1-V_{n}\left(  1-kt/n\right)  -kt/n\right]
dt+o_{p}\left(  k^{-1/2}\right)  ,
\end{align*}
and%
\begin{align*}
\Omega_{n2}  &  =\left(  \lambda-1\right)  \left(  n/k\right)  \left[
k/n-\left(  1-U_{n-k,n}\right)  \right]
{\displaystyle\int_{1/k}^{1}}
t^{\lambda-1}\log tdt\\
&  \ \ \ \ \ \ \ \ \ \ \ \ \ \ \ \ \ \ +\left(  n/k\right)  \left[
k/n-\left(  1-U_{n-k,n}\right)  \right]
{\displaystyle\int_{1/k}^{1}}
t^{\lambda-1}dt+o_{p}\left(  k^{-1/2}\right) \\
&  =\lambda^{-2}\left(  n/k\right)  \left[  k/n-\left(  1-U_{n-k,n}\right)
\right]  +o_{p}\left(  k^{-1/2}\right)  .
\end{align*}
Fix $0<\tau<1/2,$ then using approximation (\ref{approxi}), in Lemma
\ref{Lemma1}, yields%
\begin{align*}
\sqrt{k}\Omega_{n1}  &  =\left(  \lambda-1\right)  \sqrt{n/k}%
{\displaystyle\int_{1/k}^{1}}
t^{\lambda-2}\left(  \log t\right)  B_{n}\left(  1-kt/n\right)  dt\\
&  \ \ \ \ \ \ \ +\sqrt{n/k}%
{\displaystyle\int_{1/k}^{1}}
t^{\lambda-2}B_{n}\left(  1-kt/n\right)  dt+\sqrt{k}\widetilde{\Omega}%
_{n1}\left(  \tau\right)  +o_{p}\left(  1\right)  ,
\end{align*}
and%
\[
\sqrt{k}\Omega_{n2}=-\lambda^{-2}\sqrt{n/k}B_{n}\left(  1-k/n\right)
+\sqrt{k}\widetilde{\Omega}_{n2}\left(  \tau\right)  +o_{p}\left(  1\right)
,
\]
where%
\begin{align*}
\sqrt{k}\widetilde{\Omega}_{n1}\left(  \tau\right)   &  =\left(
\lambda-1\right)  O_{p}\left(  n^{-\tau}\right)  \left(  k/n\right)
^{1/2-\tau}\left(  n/k\right)  ^{1/2}%
{\displaystyle\int_{0}^{1}}
t^{\lambda-2+\left(  1/2-\tau\right)  }\left\vert \log t\right\vert dt\\
&  \ \ \ \ \ \ \ \ \ \ \ \ \ \ \ \ \ \ \ \ \ \ \ \ \ \ \ \ \ +O_{p}\left(
n^{-\tau}\right)  \left(  k/n\right)  ^{1/2-\tau}\sqrt{n/k}%
{\displaystyle\int_{0}^{1}}
t^{\lambda-2+\left(  1/2-\tau\right)  }dt,
\end{align*}
and%
\[
\sqrt{k}\widetilde{\Omega}_{n2}\left(  \tau\right)  =\lambda^{-2}O_{p}\left(
n^{-\tau}\right)  \sqrt{n/k}\left(  k/n\right)  ^{1/2-\tau}.
\]
For $\lambda>1,$ $%
{\displaystyle\int_{0}^{1}}
t^{\lambda-2+\left(  1/2-\tau\right)  }\left\vert \log t\right\vert dt=\left(
\lambda-1/2-\tau\right)  ^{-2}$ and $%
{\displaystyle\int_{0}^{1}}
t^{\lambda-2+\left(  1/2-\tau\right)  }dt=\left(  \lambda-1/2-\tau\right)
^{-1}$ are finite integrals. Then both quantities $\sqrt{k}\widetilde{\Omega
}_{n1}$ and $\sqrt{k}\widetilde{\Omega}_{n2}$ are equal to $O_{p}\left(
k^{-\tau}\right)  $ for all large $n,$ which tends in probability to $0$ as
$n\rightarrow\infty.$ Recall that up to now we have showed that%
\begin{align*}
\sqrt{k}\Omega_{n1}  &  =\left(  \lambda-1\right)  \sqrt{n/k}%
{\displaystyle\int_{1/k}^{1}}
t^{\lambda-2}\left(  \log t\right)  B_{n}\left(  1-kt/n\right)  dt\\
&  \ \ \ \ \ \ \ \ \ \ \ \ \ \ \ \ \ \ \ \ \ \ \ \ +\sqrt{n/k}%
{\displaystyle\int_{1/k}^{1}}
t^{\lambda-2}B_{n}\left(  1-kt/n\right)  dt+o_{p}\left(  1\right)  ,
\end{align*}
and
\[
\sqrt{k}\Omega_{n2}=\lambda^{-2}\sqrt{n/k}B_{n}\left(  1-k/n\right)
+o_{p}\left(  1\right)  .
\]
It remains to prove that%
\begin{align*}
&  I_{n}%
\begin{array}
[c]{c}%
:=
\end{array}
\left(  \lambda-1\right)  \sqrt{n/k}%
{\displaystyle\int_{0}^{1/k}}
t^{\lambda-2}\left(  \log t\right)  B_{n}\left(  1-kt/n\right)  dt\\
&
\ \ \ \ \ \ \ \ \ \ \ \ \ \ \ \ \ \ \ \ \ \ \ \ \ \ \ \ \ \ \ \ \ \ \ \ \ \ +\sqrt
{n/k}%
{\displaystyle\int_{0}^{1/k}}
t^{\lambda-2}B_{n}\left(  1-kt/n\right)  dt,
\end{align*}
converges, in probability, to $0.$ Indeed, since $E\left\vert B_{n}\left(
1-ks/n\right)  \right\vert \leq\sqrt{ks/n},$ then%
\[
\mathbf{E}\left\vert I_{n}\right\vert \leq\left(  \lambda-1\right)
{\displaystyle\int_{0}^{1/k}}
t^{\lambda-2+1/2}\left(  \left\vert \log t\right\vert +1\right)  dt.
\]
Since%
\[%
{\displaystyle\int_{0}^{1/k}}
t^{\lambda-2+1/2}\left(  \left\vert \log t\right\vert +1\right)  dt=\frac
{2}{\left(  2\lambda-1\right)  ^{2}}k^{-\lambda+\frac{1}{2}}\left(
2\lambda-\log k+2\lambda\log k+1\right)  ,
\]
which tends to $0$ as $n\rightarrow\infty,$ then $I_{n}$ converges to $0$ in
probability. This completes the proof of Theorem \ref{Theorem1}.\hfill
$\square$

\subsection{Proof \textbf{of Theorem} \textbf{\ref{Theorem2}.}}

\noindent To establish the asymptotic normality of $\widehat{\alpha}$, given
in $\left(  \ref{asympt-alpha}\right)  ,$ we proceed by similar arguments as
for $\widehat{\mu}_{n}$ in the proof of Theorem \ref{Theorem4}. \hfill
$\square$

\subsection{Proof \textbf{of Theorem} \textbf{\ref{Theorem3}}}

\noindent Let us divide the integral (\ref{mu}), in two parts, as follows:%
\[
\mu=\mu_{1,n}\left(  k\right)  +\mu_{2,n}\left(  k\right)  ,
\]
where%
\[
\mu_{1,n}\left(  k\right)  :=\int_{0}^{k/n}Q\left(  1-s\right)  ds\text{ and
}\mu_{2,n}\left(  k\right)  :=\int_{k/n}^{1}Q\left(  1-s\right)  ds.
\]
Recall that, in Section 1 formula (\ref{sum}), we have defined estimator
$\widehat{\mu}_{n}$ of $\mu$ by
\[
\widehat{\mu}_{n}=\widehat{\mu}_{1,n}\left(  k\right)  +\widehat{\mu}%
_{2,n}\left(  k\right)  ,
\]
where%
\[
\widehat{\mu}_{1,n}\left(  k\right)  :=\left(  k/n\right)  \left(
n\widehat{c}/k\right)  ^{1/\widehat{\alpha}}\left\{  \frac{\widehat{\alpha}%
}{\widehat{\alpha}-1}+\frac{\widehat{d}\widehat{c}^{-\widehat{\beta}%
/\widehat{\alpha}}\left(  k/n\right)  ^{\widehat{\beta}/\widehat{\alpha}-1}%
}{\widehat{\beta}-1}\right\}
\]
and
\[
\widehat{\mu}_{2,n}\left(  k\right)  :=\frac{1}{n}\sum_{i=k+1}^{n}%
X_{n-i+1,n}.
\]
To simplify notations, let us set%
\begin{equation}
Z_{ni}:=\frac{\sqrt{n}}{\sqrt{k/n}\left(  nc/k\right)  ^{1/\alpha}}\left\{
\widehat{\mu}_{i,n}\left(  k\right)  -\mu_{i,n}\left(  k\right)  \right\}
,\text{ }i=1,2. \label{twostat}%
\end{equation}
First, we consider $Z_{n1}.$ It is easy to verify that, as $n\rightarrow
\infty$%
\[
\mu_{1,n}\left(  k\right)  =\left\{  1+o_{p}\left(  1\right)  \right\}
\frac{k}{n}\left(  nc/k\right)  ^{1/\alpha}\frac{\alpha}{\alpha-1},
\]
and, under the condition (\ref{K}), we have%
\[
\widehat{\mu}_{1,n}\left(  k\right)  =\left\{  1+o_{p}\left(  1\right)
\right\}  \frac{k}{n}\frac{\widehat{\alpha}}{\widehat{\alpha}-1}\left(
n\widehat{c}/k\right)  ^{1/\widehat{\alpha}}.
\]
It follows that
\[
\widehat{\mu}_{1,n}\left(  k\right)  -\mu_{1,n}\left(  k\right)  =\left\{
1+o_{p}\left(  1\right)  \right\}  \frac{k}{n}\left\{  \frac{\widehat{\alpha}%
}{\widehat{\alpha}-1}\left(  n\widehat{c}/k\right)  ^{1/\widehat{\alpha}%
}-\frac{\alpha}{\alpha-1}\left(  nc/k\right)  ^{1/\alpha}\right\}  .
\]
Let us write $Z_{n1}=T_{1n}+T_{2n},$ where%
\[
T_{1n}:=\left\{  1+o_{p}\left(  1\right)  \right\}  \sqrt{k}\left\{
\frac{\widehat{\alpha}}{\widehat{\alpha}-1}-\frac{\alpha}{\alpha-1}\right\}
,
\]
and%
\[
T_{2n}:=\left\{  1+o_{p}\left(  1\right)  \right\}  \sqrt{k}\left\{
\frac{\left(  n\widehat{c}/k\right)  ^{1/\widehat{\alpha}}}{\left(
nc/k\right)  ^{1/\alpha}}-1\right\}  .
\]
We begin by showing that $T_{2n}\overset{P}{\rightarrow}0,$ as $n\rightarrow
\infty.$ First observe that $T_{2n}$ may be rewritten into
\[
T_{2n}=\left\{  1+o_{p}\left(  1\right)  \right\}  \sqrt{k}\left\{  \left(
nc/k\right)  ^{1/\widehat{\alpha}-1/\alpha}\left(  \widehat{c}/c\right)
^{1/\widehat{\alpha}-1/\alpha}-1\right\}  .
\]
Assumptions $\left(  i\right)  $ and $\left(  ii\right)  $ of Theorem
\ref{Theorem1} imply that $k^{1/2}/\log\left(  n/k\right)  \rightarrow
\infty.\ $Also, from Theorem 1 of \cite{PeQi04}, the asymptotic normality of
$\widehat{\alpha}$ gives $\widehat{\alpha}-\alpha=O_{p}\left(  k^{-1/2}%
\right)  .$ Therefore $\left(  1/\widehat{\alpha}-1/\alpha\right)  \log\left(
nc/k\right)  \overset{P}{\rightarrow}0,$ this implies that $\left(
nc/k\right)  ^{1/\widehat{\alpha}-1/\alpha}\overset{P}{\rightarrow}1,$ as
$n\rightarrow\infty.$ On the other hand, from equation $\left(  4.7\right)  $
in \cite{LiPeYa10}, we have
\[
\widehat{c}/c-1=\alpha^{-1}\left(  \widehat{\alpha}-\alpha\right)  \log
\dfrac{n}{k}+o_{p}\left(  k^{-1/2}\log\dfrac{n}{k}\right)  .
\]
Since $\widehat{c}$ is a consistent estimator of $c,$ then Taylor's expansion
gives
\[
\left(  \widehat{c}/c\right)  ^{1/\widehat{\alpha}-1/\alpha}-1=\alpha
^{-1}\left(  1+o_{p}\left(  1\right)  \right)  \left(  \widehat{\alpha}%
-\alpha\right)  \left(  \widehat{c}/c-1\right)  ,\text{ as }n\rightarrow
\infty.
\]
It suffices now to show that $\sqrt{k}\left(  \left(  \widehat{c}/c\right)
^{1/\widehat{\alpha}-1/\alpha}-1\right)  $ converges to $0$ in
probability.$\ $Indeed, again by using the fact that $\widehat{\alpha}%
-\alpha=O_{p}\left(  k^{-1/2}\right)  ,$ yields%
\[
\sqrt{k}\left(  \left(  \widehat{c}/c\right)  ^{1/\widehat{\alpha}-1/\alpha
}-1\right)  =O_{p}\left(  1\right)  \left(  k^{-1/2}\log\dfrac{n}{k}%
+o_{p}\left(  \frac{\log\dfrac{n}{k}}{\sqrt{k}}\right)  \right)  ,
\]
which tends in probability to $0,$ because we already have $\sqrt{k}%
/\log\left(  n/k\right)  \rightarrow\infty.$ Now, we consider the term
$T_{1n}.$ Since $\widehat{\alpha}$ is a consistent estimator of $\alpha,$ then
it is easy to show that%
\[
T_{1n}=-\frac{1+o_{p}\left(  1\right)  }{\left(  \alpha-1\right)  ^{2}}%
\sqrt{k}\left(  \widehat{\alpha}-\alpha\right)  ,\text{ as }n\rightarrow
\infty.
\]
From \ref{Theorem1}, we infer that
\[
T_{1n}=-\frac{\left(  1+o_{p}\left(  1\right)  \right)  \alpha}{\left(
\alpha-1\right)  ^{2}}\left(  \eta_{1}W_{1n}+\eta_{2}W_{2n}+\eta_{3}%
W_{3n}\right)  ,\text{ as }n\rightarrow\infty.
\]
It follows that
\begin{equation}
Z_{n1}=-\frac{\alpha}{\left(  \alpha-1\right)  ^{2}}\left\{  \eta_{1}%
W_{1n}+\eta_{2}W_{2n}+\eta_{3}W_{3n}\right\}  +o_{p}\left(  1\right)  .
\label{Zn1}%
\end{equation}
Let us now consider the asymptotic distribution of $Z_{n2},$ in (\ref{twostat}%
). It is shown in \cite{CsMa85} or more recently in \cite{NeMe09} that%
\[
Z_{n2}=-\frac{\int_{k/n}^{1}B_{n}\left(  1-s\right)  dQ\left(  1-s\right)
}{\sqrt{k/n}Q\left(  1-k/n\right)  }+o_{p}\left(  1\right)  .
\]
On the other hand, from (\ref{A3}), we have $Q\left(  1-k/n\right)
\sim\left(  nc/k\right)  ^{1/\alpha},$ as $n\rightarrow\infty,$ it follows
that%
\begin{equation}
Z_{n2}=W_{4n}+o_{p}\left(  1\right)  . \label{Zn2}%
\end{equation}
Combining (\ref{Zn1}) and (\ref{Zn2}) achieves the proof of Theorem
\ref{Theorem3}.\hfill$\square$

\subsection{Proof \textbf{of Theorem} \textbf{\ref{Theorem4}.}}

\noindent Now, we investigate the asymptotic normality of $\widehat{\mu}_{n}$
given in$\left(  \ref{asympt-mu}\right)  $.\ Since $W_{in},$ $i=1,...,4$ are
sequences of centred Gaussian r.v.'s, then%
\[
\frac{\sqrt{n}}{\sqrt{k/n}\left(  nc/k\right)  ^{1/\alpha}}\left\{
\widehat{\mu}_{n}-\mu\right\}  \overset{d}{\rightarrow}\mathcal{N}\left(
0,\Gamma\Sigma\Gamma^{t}\right)  ,\text{ as }n\rightarrow\infty,
\]
where
\[
\Gamma:=\left(  -\frac{\alpha}{\left(  \alpha-1\right)  ^{2}}\eta_{1}%
,-\frac{\alpha}{\left(  \alpha-1\right)  ^{2}}\eta_{2},-\frac{\alpha}{\left(
\alpha-1\right)  ^{2}}\eta_{3},1\right)  ,
\]
$\Gamma^{t}$ is the transpose of $\Gamma$ and $\Sigma$ is the
variance-covariance matrix of the vector $\left(  W_{1n},...,W_{4n}\right)  $
defined by%
\[
\sum=\left[
\begin{array}
[c]{cccc}%
1 & \frac{1}{\lambda^{2}}-\frac{1}{\lambda} & \frac{2}{\lambda^{3}}-\frac
{1}{\lambda^{2}} & -1\\
\frac{1}{\lambda^{2}}-\frac{1}{\lambda} & \frac{1}{2\lambda-1}-\frac
{1}{\lambda^{2}} & \frac{1}{\left(  2\lambda-1\right)  ^{2}}-\frac{1}%
{\lambda^{3}} & \frac{\lambda-1}{\lambda}\\
\frac{2}{\lambda^{3}}-\frac{1}{\lambda^{2}} & \frac{1}{\left(  2\lambda
-1\right)  ^{2}}-\frac{1}{\lambda^{3}} & \frac{2}{\left(  2\lambda-1\right)
^{3}}-\frac{1}{\lambda^{4}} & -\frac{1}{\lambda^{2}}\\
-1 & \frac{\lambda-1}{\lambda} & -\frac{1}{\lambda^{2}} & \frac{2}{2-\alpha}%
\end{array}
\right]  .
\]
Note that the elements of $\sum$ were obtained after tedious computations of
the limits of the expectations $\mathbf{E}\left[  W_{in}W_{jn}\right]  $ for
$i,j=1,4$ $\left(  i\leq j\right)  $.\ Analogue calculus of these quantities
may be found in \cite{Pe01}, \cite{NeMe09} and \cite{NeMe10}. Finally, a
standard calculation of the product $\Gamma\Sigma\Gamma^{t}$ yields%
\[
\Gamma\Sigma\Gamma^{t}=\frac{\alpha^{2}\beta^{4}}{\left(  \alpha-1\right)
^{4}\left(  \alpha-\beta\right)  ^{4}}+\frac{2}{2-\alpha}+\frac{2\alpha
\beta^{2}}{\left(  \alpha-1\right)  ^{2}\left(  \alpha-\beta\right)  ^{2}},
\]
which is denoted by $\sigma^{2}\left(  \alpha,\beta\right)  .$ This completes
the proof of Theorem \ref{Theorem4}\textbf{.}\hfill$\square$

\section{\textbf{Concluding notes\label{concluding notess}}}

\noindent The main objective of this paper was to propose a bias-reduced
estimator for the mean of a heavy-tailed distribution. This was achieved on
the basis of the bias-reduction of the first and second order parameter
estimators of regularly varying distributions developed by \cite{PeQi04} and
the corresponding high quantiles estimators introduced by \cite{LiPeYa10}. In
addition, the newly introduced estimator is asymptotically normal, making
confidence intervals easily constructible. We conclude by simulation that,
compared to that of Peng, our new estimator has smaller bias and RMSE and
consequently it performs better.

\appendix

\section{Appendix}

\subsection{Auxiliary results}

\begin{lemma}
\label{Lemma2}Let $k=k_{n}$ be a sequence of integers satisfying (\ref{K}) and
$f\left(  x\right)  =x^{\lambda-1}\log x,$ $\lambda>1.$ Then,\ uniformly on
$s\in\left[  1/n,k/n\right]  ,$ we have%
\[
f^{\prime}\left(  \dfrac{\varphi_{n}\left(  s\right)  }{1-U_{n-k,n}}\right)
-f^{\prime}\left(  \dfrac{s}{1-U_{n-k,n}}\right)  =o_{p}\left(  1\right)
\left(  \dfrac{s}{k/n}\right)  ^{\lambda-2}\log\dfrac{s}{k/n},\text{ as
}n\rightarrow\infty.
\]

\end{lemma}

\noindent\textbf{Proof.} We have $n\left(  1-U_{n-k,n}\right)  /k\overset
{p}{\rightarrow}0,$ as $n\rightarrow\infty,$ then
\[
f^{\prime}\left(  \dfrac{\varphi_{n}\left(  s\right)  }{1-U_{n-k,n}}\right)
-f^{\prime}\left(  \dfrac{s}{1-U_{n-k,n}}\right)  =\left(  1+o_{p}\left(
1\right)  \right)  \left[  f^{\prime}\left(  \dfrac{\varphi_{n}\left(
s\right)  }{k/n}\right)  -f^{\prime}\left(  \dfrac{s}{k/n}\right)  \right]  .
\]
A straightforward calculation of the derivative of $f$ yields%
\begin{align}
f^{\prime}\left(  \dfrac{\varphi_{n}\left(  s\right)  }{k/n}\right)
-f^{\prime}\left(  \dfrac{s}{k/n}\right)   &  =\left(  \dfrac{s}{k/n}\right)
^{\lambda-2}\left[  \left(  \lambda-1\right)  \left(  \left(  \dfrac
{\varphi_{n}\left(  s\right)  }{s}\right)  ^{\lambda-2}-1\right)  \log
\dfrac{\varphi_{n}\left(  s\right)  }{k/n}\right. \nonumber\\
&  +\left.  \left(  \lambda-1\right)  \log\dfrac{\varphi_{n}\left(  s\right)
}{s}+\left(  \dfrac{\varphi_{n}\left(  s\right)  }{s}\right)  ^{\lambda
-2}-1\right]  . \label{eql}%
\end{align}
Observe now, that inequalities (\ref{phi}) imply%
\[
\min\left\{  \frac{1-s-V_{n}\left(  1-s\right)  }{s},0\right\}  <\dfrac
{\varphi_{n}\left(  s\right)  }{s}-1<\max\left\{  \frac{1-s-V_{n}\left(
1-s\right)  }{s},0\right\}  .
\]
From \cite{Wel78}, we have%
\[
\sup_{1/n\leq s<1}\frac{\left\vert 1-s-V_{n}\left(  1-s\right)  \right\vert
}{s}\overset{p}{\rightarrow}0\text{ as }n\rightarrow\infty,
\]
it follows that%
\begin{equation}
\sup_{1/n\leq s\leq k/n}\left\vert \dfrac{\varphi_{n}\left(  s\right)  }%
{s}-1\right\vert \overset{p}{\rightarrow}0\text{ as }n\rightarrow\infty.
\label{sup2}%
\end{equation}
On the other hand, in view Lemma 3 in \cite{NeMe09}, we infer that%
\begin{equation}
\sup_{1/n\leq s\leq k/n}\left\vert \dfrac{s}{\varphi_{n}\left(  s\right)
}-1\right\vert \overset{p}{\rightarrow}0\text{ as }n\rightarrow\infty.
\label{sup1}%
\end{equation}
By applying the mean value theorem to the functions $x\rightarrow\log x$ and
$x\rightarrow x^{\lambda-1}$ respectively, then by using (\ref{sup2}) and
(\ref{sup1}), we show readily that, as $n\rightarrow\infty$%
\begin{equation}
\sup_{1/n\leq s\leq k/n}\left\vert \log\dfrac{\varphi_{n}\left(  s\right)
}{s}\right\vert \overset{p}{\rightarrow}0\text{ and }\sup_{1/n\leq s\leq
k/n}\left\vert \left(  \dfrac{\varphi_{n}\left(  s\right)  }{s}\right)
^{\lambda-1}-1\right\vert \overset{p}{\rightarrow}0. \label{T1}%
\end{equation}
Note that the first result of (\ref{T1}) implies that
\begin{equation}
\sup_{1/n\leq s\leq k/n}\left\vert \log\dfrac{\varphi_{n}\left(  s\right)
}{k/n}-\log\dfrac{s}{k/n}\right\vert \overset{p}{\rightarrow}0,\text{ as
}n\rightarrow\infty. \label{T2}%
\end{equation}
By using equations (\ref{T1}) and (\ref{T2}) together, we show that, uniformly
in $s\in\left[  1/n,k/n\right]  ,$ the right-hand side of equation $\left(
\ref{eql}\right)  ,$ is equal to $o_{p}\left(  1\right)  \left(  \dfrac
{s}{k/n}\right)  ^{\lambda-2}\log\dfrac{s}{k/n}.$\hfill$\square$

\begin{lemma}
\label{Lemma3} We have $\sqrt{k}\Omega_{n1}^{\ast}\overset{p}{\rightarrow}0$
and $\sqrt{k}\Omega_{n2}^{\ast}\overset{p}{\rightarrow}0,$ as $n\rightarrow
\infty.$
\end{lemma}

\noindent Proof. We only show the first result. The second one is obtained by
similar arguments. Recall that%
\begin{align*}
\Omega_{n1}^{\ast}  &  :=-\left(  n/k\right)
{\displaystyle\int_{1/n}^{k/n}}
\left[  f^{\prime}\left(  \dfrac{\varphi_{n}\left(  s\right)  }{1-U_{n-k,n}%
}\right)  -f^{\prime}\left(  \dfrac{s}{1-U_{n-k,n}}\right)  \right] \\
&  \ \ \ \ \ \ \ \ \ \ \ \ \ \ \ \ \ \ \ \ \ \ \ \ \ \ \ \ \ \ \ \times\left[
\dfrac{1-V_{n}\left(  1-s\right)  }{1-U_{n-k,n}}-\dfrac{s}{1-U_{n-k,n}%
}\right]  ds.
\end{align*}
Using Lemma \ref{Lemma2}, we get, as $n\rightarrow\infty$%
\[
\Omega_{n1}^{\ast}=o_{p}\left(  1\right)  \left(  n/k\right)
{\displaystyle\int_{1/n}^{k/n}}
\left(  \dfrac{s}{k/n}\right)  ^{\lambda-2}\left\vert \log\dfrac{s}%
{k/n}\right\vert \left\vert \dfrac{1-V_{n}\left(  1-s\right)  }{k/n}-\dfrac
{s}{k/n}\right\vert ds.
\]
By a change of variables, we get%
\[
\Omega_{n1}^{\ast}=o_{p}\left(  1\right)  \left(  n/k\right)
{\displaystyle\int_{1/k}^{1}}
s^{\lambda-2}\left\vert \log s\right\vert \left\vert 1-V_{n}\left(
1-ks/n\right)  -ks/n\right\vert ds.
\]
Making use of approximation$\left(  \ref{approxi}\right)  $ yields%
\[
\sqrt{k}\Omega_{n1}^{\ast}=o_{p}\left(  1\right)  \left(  n/k\right)  ^{1/2}%
{\displaystyle\int_{1/k}^{1}}
s^{\lambda-2}\left\vert \log s\right\vert \left(  \left\vert B_{n}\left(
1-ks/n\right)  \right\vert +\left(  ks/n\right)  ^{1/2-\tau}O_{p}\left(
n^{-\tau}\right)  \right)  ds.
\]
In other words%
\begin{align*}
\sqrt{k}\Omega_{n1}^{\ast}  &  =o_{p}\left(  1\right)  \left(  n/k\right)
^{1/2}%
{\displaystyle\int_{1/k}^{1}}
s^{\lambda-2}\left\vert \log s\right\vert \left\vert B_{n}\left(
1-ks/n\right)  \right\vert ds.\\
&
\ \ \ \ \ \ \ \ \ \ \ \ \ \ \ \ \ \ \ \ \ \ \ \ \ \ \ \ \ \ \ \ \ \ \ +o_{p}%
\left(  k^{-\tau}\right)
{\displaystyle\int_{1/k}^{1}}
s^{\lambda-3/2-\tau}\left\vert \log s\right\vert ds.
\end{align*}
The expectation of the first term of right-hand side of the previous equation
is less than or equal to%
\[
o_{p}\left(  1\right)  \left(  n/k\right)  ^{1/2}%
{\displaystyle\int_{0}^{1}}
s^{\lambda-2}\left\vert \log s\right\vert \mathbf{E}\left[  \left\vert
B_{n}\left(  1-ks/n\right)  \right\vert \right]  ds.
\]
Using the fact that $E\left\vert B_{n}\left(  1-ks/n\right)  \right\vert
\leq\left(  ks/n\right)  ^{1/2},$ we show that the previous quantity is less
than or equal to $o_{p}\left(  1\right)
{\displaystyle\int_{0}^{1}}
s^{\lambda-3/2}\left\vert \log s\right\vert ds.$ Since both integrals $%
{\displaystyle\int_{0}^{1}}
s^{\lambda-3/2-\tau}\left\vert \log s\right\vert ds$ and $%
{\displaystyle\int_{0}^{1}}
s^{\lambda-3/2-\tau}\left\vert \log s\right\vert ds$ are finite, then
$\sqrt{k}\Omega_{n1}^{\ast}=o_{p}\left(  1\right)  ,$ as $n\rightarrow
\infty.\hfill\square$

\subsection{Optimal choice of the sample fraction $k$\label{RT}}

\noindent Reiss and Thomas in \cite[, page 137]{ReTo07}, proposed a heuristic
method for choosing the optimal number of upper extremes used in the
computation of the tail index estimate. In this paper, we adopt this algorithm
by making use of Peng and Qi estimator $\widehat{\alpha}=\widehat{\alpha
}\left(  k\right)  $ which is defined by the system of two equations
(\ref{alpha-beta}). By this methodology, one defines the optimal sample
fraction of upper order statistics $k^{\ast}$ by%
\[
k^{\ast}:=\arg\min_{k}\frac{1}{k}\sum_{i=1}^{k}i^{\theta}\left\vert
\widehat{\alpha}\left(  i\right)  -\text{median}\left\{  \widehat{\alpha
}\left(  1\right)  ,...,\widehat{\alpha}\left(  k\right)  \right\}
\right\vert ,
\]
with suitable constant $0\leq\theta\leq1/2.$ The quantity $\widehat{\alpha
}\left(  i\right)  $ corresponds to Peng and Qi estimator of the shape
parameter $\alpha,$ based on the $i$ upper order statistics.$\ $On the light
of our simulation study,\textbf{\ }we obtained reasonable results by choosing
$\theta=0.3.$ The same value for $\theta$ has also been observed by
\cite{NaFr04} when employing Hill's estimator.\ The software programs of this
methodology are incorporated in the "Xtremes" package accompanying the book of
\cite{ReTo07}.\bigskip

{\small \noindent}\textbf{E-mail addresses:}

{\small \noindent\texttt{brah.brahim@gmail.com} (B.~Brahimi),}

{\small \noindent\texttt{djmeraghni@yahoo.com} (D.~Meraghni),}

{\small \noindent\texttt{yahia\_dj@yahoo.fr} (D.~Yahia). }


\begin{thebibliography}{9999999999999999999999999999999999999}                                                            %


\bibitem[Beirlant \textit{et al.} (2001)]{BeMaDi01}Beirlant, J., Matthys, G.,
Diercks, G., 2001. Heavy-tailed distributions and rating. Astin Bull. 31, no.
1, 37-58.

\bibitem[Beirlant \textit{et al. }(2002)]{BeDiSt02}Beirlant, J., Diercks, G.,
Guillou, A., St\u{a}ric\u{a}, C., 2002. On exponential representations of
log-spacings of extreme order statistics. Extremes 5, no. 2, 157-180.

\bibitem[Beirlant \textit{et al. }(2008)]{BeFiGV08}Beirlant, J., Figueiredo,
F., Gomes, M.I., Vandewalle, B., 2008. Improved reduced-bias tail index and
quantile estimators. J. Statist. Plann. Inference 138, no. 6, 1851-1870.

\bibitem[Brahimi \textit{et al.} (2011)]{BrMeNe11}Brahimi, B., Meraghni, D.,
Necir, A., Zitikis, R., 2011. Estimating the distortion parameter of the
proportional hazard premium for heavy-tailed losses. Insurance: Mathematics
and Economics 49, 325-334.

\bibitem[Caeiro \textit{et al. }(2004)]{CaFiGo04}Caeiro, F., Figueiredo, F.,
Gomes, M. I., 2004. Bias reduction of a tail index estimator through an
external estimation of the second order parameter. Statistics. 38 (6), 497--510.

\bibitem[Caeiro \textit{et al. }(2009)]{CaGoRo09}Caeiro, F., Gomes, M.I.,
Rodrigues, L.H., 2009. Reduced-bias tail index estimators under a third-order
framework. Comm. Statist. Theory Methods 38, no. 6-7, 1019-1040.

\bibitem[Cheng and Peng, (2001)]{Ch Pe01}Cheng, S., Peng, L., 2001. Confidence
intervals for the tail index. Bernoulli 7, no. 5, 751-760.

\bibitem[Cs\"{o}rg\H{o} and Mason, (1985) ]{CsMa85}Cs\"{o}rg\H{o}, S., Mason,
D.M., 1985. Central limit theorems for sums of extreme values. \textit{Math.
Proc. Cambridge Philos. Soc.} 98, no. 3, 547-558.

\bibitem[Cs\"{o}rg\H{o} \textit{et al. }(1985)]{CsDeMa85}Cs\"{o}rg\H{o}, S.,
Deheuvels, P, Mason, D.M., 1985. Kernel estimates of the tail index of a
distribution. Ann. Statist. 13, no. 3, 1050--1077.

\bibitem[Cs\"{o}rg\H{o} \textit{et al.} (1986)]{CsCsHM86}Cs\"{o}rg\H{o}, M.,
Cs\"{o}rg\H{o}, S., Horv\'{a}th, L., Mason, D.M., 1986. Weighted empirical and
quantile processes. Ann. Probab. 14, no. 1, 31-85.

\bibitem[Danielsson \textit{et al. }(2001)]{DadePd01}Danielsson, J., de Haan,
L., Peng, L., de Vries, C.G., 2001. Using a bootstrap method to choose the
sample fraction in tail index estimation. J. Multivariate Anal. 76, no. 2, 226-248.

\bibitem[Dekkers and de Haan, (1993)]{DedeH93}Dekkers, A.L.M., de Haan, L.,
1993. Optimal choice of sample fraction in extreme-value estimation. J.
Multivariate Anal. 47, no. 2, 173-195.

\bibitem[Drees and Kaufmann, (1998)]{DrKa98}Drees, H., Kaufmann, E., 1998.
Selecting the optimal sample fraction in univariate extreme value estimation.
Stochastic Process. Appl. 75, no. 2, 149-172.

\bibitem[Feureverger and Hall, (1999)]{FeHa99}Feureverger, A., Hall, P., 1999.
Estimating a tail exponent by modelling departure from a Pareto distribution.
Ann. Statist. 27, no. 2, 760-781.

\bibitem[Fraga Alves \textit{et al. }(2007)]{FrGodeN07}Fraga Alves, M.I.,
Gomes, M.I., de Haan, L., Neves, C., 2007. A note on second order conditions
in extreme value theory: linking general and heavy tail conditions. REVSTAT.
5, no. 3, 285-304.

\bibitem[Goegebeur and de Wet, (2011)]{GdeW11}Goegebeur, Y., de Wet, T., 2011.
Estimation of the third-order parameter in extreme value statistics. Test DOI: 10.1007/s11749-011-0246-2.

\bibitem[Gomes and Martins, (2002)]{GoMa02}Gomes, M.I., Martins, M.J., 2002.
"Asymptotically unbiased" estimators of the tail index based on external
estimation of the second order parameter. Extremes. 5, no. 1, 5-31.

\bibitem[Gomes and Martins, (2004)]{GoMa04}Gomes, M.I., Martins, M.J., 2004.
Bias reduction and explicit semi-parametric estimation of the tail index. J.
Statist. Plann. Inference. 124, no. 2, 361-378.

\bibitem[Gomes and Figueiredo, (2006)]{GoFi06}Gomes, M.I., Figueiredo, F.,
2006. Bias reduction in risk modelling: semi-parametric quantile estimation.
Test 15, no. 2, 375-396.

\bibitem[Gomes and Pestana, (2007)]{GP07}Gomes, M.I., Pestana, D., 2007. A
sturdy reduced-bias extreme quantile (VaR) estimator. J. Amer. Statist. Assoc.
102, no. 477, 280-292.

\bibitem[de Haan and Stadtm\"{u}ller, (1996)]{deHSt96}de Haan, L.,
Stadtm\"{u}ller, U., 1996. Generalized regular variation of second order. J.
Austral. Math. Soc. Ser. A 61, no. 3, 381-395.

\bibitem[de Haan and Peng, (1998)]{deHP98}de Haan, L., Peng, L., 1998.
Comparison of tail index estimators. Statist. Neerlandica. 52, no. 1, 60-70.

\bibitem[de Haan and Ferreria, (2006)]{deHFe06}de Haan, L., Ferreria, A.,
2006. Extreme Values Theory: An introduction. Springer.

\bibitem[Hall, (1982)]{Ha82}Hall, P. 1982. On some simple estimators of an
exponent of regular variation. J. Roy. Statist. Soc. Ser. B. 44, 37-42.

\bibitem[Hall and Welsh, (1985)]{HaWe85}Hall, P., Welsh, A. H., 1985. Adaptive
estimates of parameters of regular variation. Ann. Statist. 13, 331-341.

\bibitem[Hill, (1975)]{Hill75}Hill, B.M., 1975. A simple general approach to
inference about the tail of a distribution. Ann. Statist. 3, no. 5, 1163-1174.

\bibitem[Li \textit{et al. }(2010)]{LiPeYa10}Li, D., Peng, L., Yang, J., 2010.
Bias reduction for high quantiles. J. Statist. Plann. Inference 140, no. 9, 2433-2441.

\bibitem[Mason, (1982)]{Ma82}Mason, D., 1982. Laws of large numbers for sums
of extreme values. Ann. Probab. 10, no. 3, 754-764.

\bibitem[Matthys \textit{et al }(2004)]{MaDeGuBe04}Matthys, G., Delafosse, E.,
Guillou, A., Beirlant, J., 2004. Estimating catastrophic quantile levels for
heavy-tailed distributions. Insurance Math. Econom. 34, no. 3, 517-537.

\bibitem[Necir and Meraghni, (2009)]{NeMe09}Necir, A., Meraghni, D., 2009.
Empirical estimation of the proportional hazard premium for heavy-tailed claim
amounts. Insurance Math. Econom. 45, no. 1, 49-58.

\bibitem[Necir and Meraghni, (2010)]{NeMe10}Necir, A., Meraghni, D., 2010.
Estimating L-functionals for heavy-tailed distributions and applications.
Journal of Probability and Statistics 2010, ID 707146.

\bibitem[Neves and Fraga Alves, (2004)]{NaFr04}Neves, C., Fraga Alves, M.I.,
2004. Reiss and Thomas' automatic selection of the number of extremes. Comput.
Statist. Data Anal. 47, no. 4, 689-704.

\bibitem[Peng, (2001)]{Pe01}Peng, L., 2001. Estimating the mean of a heavy
tailed distribution. Statist. Probab. Lett. 52, no. 3, 255--264.

\bibitem[Peng and Qi, (2004)]{PeQi04}Peng, L., Qi, Y., 2004. Estimating the
first- and second-order parameters of a heavy-tailed distribution. Aust. N. Z.
J. Stat. 46, no. 2, 305--312.

\bibitem[Reiss and Thomas, (2007)]{ReTo07}Reiss, R.-D., Thomas, M., 2007.
Statistical Analysis of Extreme Values with Applications to Insurance,
Finance, Hydrology and Other Fields, 3rd ed. Birkh\"{a}user Verlag, Basel,
Boston, Berlin.

\bibitem[Rolski \textit{et al.} (1999)]{RoScSc99}Rolski, T., Schimidli, H.,
Schimd, V., Teugels, J.L., 1999. Stochastic Processes for Insurance and
Finance. John Wiley and Sons, New York.

\bibitem[Weissman, (1978)]{Wei78}Weissman, I., 1978. Estimation of parameters
and large quantiles based on the $k$ largest observations. J. Amer. Statist.
Assoc. 73, no. 364, 812-815.

\bibitem[Wellner, (1978)]{Wel78}Wellner, J.A., 1978. Limit theorems for the
ratio of the empirical distribution function to the true distribution
function. Z. Wahrsch. Verw. Gebiete 45, no. 1, 73-88.
\end{thebibliography}
\end{document}